\documentclass[conference]{IEEEtran}
\IEEEoverridecommandlockouts
\usepackage{cite}
\usepackage{amsmath,amssymb,amsfonts}
\usepackage{algorithmic}
\usepackage{graphicx}
\usepackage{subcaption}
\usepackage{textcomp}
\usepackage{xcolor}
\def\BibTeX{{\rm B\kern-.05em{\sc i\kern-.025em b}\kern-.08em
    T\kern-.1667em\lower.7ex\hbox{E}\kern-.125emX}}
\begin{document}

\title{On the Nonlinear Sensitivity of Phononic Frequency Combs to Physical Perturbations}

\author{\IEEEauthorblockN{Madhurendra Mishra}
\IEEEauthorblockA{\textit{Department of Physics} \\
\textit{Sri Guru Tegh Bahadur Khalsa College}\\
\textit{University of Delhi}\\
Delhi, India \\
madhurendramishra24@gmail.com}
\and
\IEEEauthorblockN{Zhen Qi}
\IEEEauthorblockA{\textit{Department of Biomedical Engineering}\\
\textit{Worcester Polytechnic Institute}\\
Worcester, MA 01609, United States \\zqi@wpi.edu
}
\and
\IEEEauthorblockN{Adarsh Ganesan}
\IEEEauthorblockA{\textit{Department of Electrical and} \\
\textit{Electronics Engineering}\\
\textit{BITS Pilani, Dubai Campus}\\
Dubai, UAE\\
adarsh@dubai.bits-pilani.ac.in}
}

\maketitle

\begin{abstract}
Phononic frequency combs offer a rich platform for nonlinear sensing, yet how their observable properties respond to changes in physical parameters remains poorly understood. Using a reduced two-mode autoparametric resonance model, we investigate how primary and secondary detuning, drive amplitude, and relative damping jointly shape amplitude and frequency sensitivity across the nonlinear parameter space. We find that sensitivity is far from uniform: primary detuning shifts the comb response smoothly, secondary detuning produces sharply localized transitions near resonance manifolds, and drive amplitude concentrates peak sensitivity close to the activation threshold rather than deep within the comb state. The relative damping redistributes energy continuously between modes without introducing discontinuities. The nonlinear sensitivity of amplitude and frequency observables across all parameters points to a common physical origin in autoparametric resonance, nonlinear saturation, and 
coupling-induced synchronization, offering a coherent basis for designing nonlinear sensing platforms with deliberate, parameter-aware sensitivity engineering.
\end{abstract}


\section{Introduction}

Micro- and nano-electromechanical systems (MEMS/NEMS) have long served as a productive arena for exploring nonlinear dynamics. Driven beyond the linear regime, mechanical resonators reveal rich internal structure--mode coupling, bifurcations, and energy redistribution--that reflects the deeper geometry of the underlying dynamical landscape. Among the most consequential manifestations of such dynamics is the phononic frequency comb (PFC): a coherent, discretely spaced array of spectral lines that emerges
when vibrational modes interact parametrically.

The frequency comb concept originates in optics \cite{b1,b2}, where it transformed precision metrology by linking optical and microwave frequencies with unprecedented accuracy. In both time and frequency domains, a comb carries a clear physical signature--periodic pulse trains and equidistant spectral lines--whose coherence reflects phase-locked oscillations. Extending this coherence to mechanical systems was non-trivial, given the dissipation and disorder endemic to micro-scale devices, and its eventual demonstration marked a genuine conceptual broadening of comb physics.

PFCs emerge from nonlinear intermodal coupling, wherein energy pumped into one vibrational mode is redistributed into adjacent modes through three- or four-wave mixing. The first experimental realization was achieved through intrinsic three-wave mixing in a micromechanical resonator by Ganesan et al. \cite{b3}, consistent with earlier theoretical predictions \cite{b4}. The autoparametric mechanism selectively amplifies
sidebands when internal resonance conditions are met, producing a cascade of phase-coherent lines with uniform spacing. Since that demonstration, PFCs have been observed across a wide range of platforms: silicon beam resonators \cite{b5}, two-dimensional material systems \cite{b6}, quartz resonators \cite{b7}, rhombic, arch, and ring micro-resonator geometries \cite{b8,b9,b10}, and optomechanical, magnetomechanical, and electromechanical systems \cite{b11,b12,b13}. Theoretical work has further extended comb formation to molecular systems \cite{b14}, twisted bilayer van der Waals materials \cite{b15}, and solid-state platforms \cite{b16}, reinforcing the view that wherever nonlinearity and modal coupling coexist, comb formation is latent.

Beyond their fundamental interest, PFCs offer practical advantages for sensing. Multiple phase-coherent spectral lines allow information to be encoded and read out across the entire comb, providing frequency-domain resolution and redundancy unavailable in single-mode resonators. Comb line spacing can be engineered to yield ultrasensitive temperature responses \cite{b17}, and applications have been demonstrated in underwater imaging and acoustic communication \cite{b18,b19}, vibration energy harvesting \cite{b20,b21}, and a portfolio of patented technologies \cite{b22,b23,b24,b25,b26,b27}.

A significant theoretical advance was the derivation of existence conditions for PFCs in coupled-mode systems \cite{b28}, establishing the parameter regimes in which stable comb formation occurs. Alongside this, frameworks for engineering MEMS/NEMS devices with tunable comb sensitivity have been proposed \cite{b29}. These results identify where in parameter space a comb can exist, but leave open the equally important question of how comb observables respond once it does. In any real device, parameters drift and bias voltages are tuned; the sensitivity of the comb to such variations is a prerequisite for principled design, not a secondary concern.

Several recent studies have begun to address this. A drive-controlled transition from negative to positive parametric dependence has been demonstrated within a van der Pol--Mathieu framework \cite{b30}, showing that the sign of sensitivity is itself experimentally controllable. In the radiofrequency domain, passive comb generation through nonlinear interactions in piezoelectric microacoustic resonators has enabled battery-free ranging in GPS-denied environments \cite{b31}--a striking illustration of how parametric physics finds new applications across frequency regimes. The relevance of PFC dynamics to quantum sensing has also sharpened: quantum squeezing
demonstrated in a nonlinear mechanical oscillator \cite{b32} establishes a direct bridge between classical comb formation and quantum-limited precision. In our own prior work \cite{b33}, we examined how PFC amplitudes and frequencies evolve under systematic detuning variations, mapping the qualitative landscape of amplitude-frequency tunability in a nonlinear sensing context.

Despite substantial progress in establishing the conditions for PFC formation, a systematic and quantitative understanding of how comb observables respond to parameter perturbations across the full nonlinear parameter space remains lacking. In strongly nonlinear coupled systems, such responses are inherently non-uniform: a single parameter perturbation may induce smooth variations in some observables while triggering abrupt transitions, discontinuities, or sign reversals in others. Sensitivity may localize near activation thresholds, collapse under nonlinear saturation, or concentrate along narrow resonance manifolds. Yet no unified framework exists that captures how measurable quantities--modal intensities, peak-to-peak amplitudes, and oscillation frequencies--respond differentially to variations in system parameters, nor has the role of nonlinear coupling strength $\gamma_{21}$ in shaping this landscape been seriously addressed. This gap limits both theoretical insight and the rational design of comb-based sensing architectures.

In this work, we address this directly by systematically mapping amplitude- and frequency-based observables across the full nonlinear parameter space $(\Delta_1, \Delta_2, f, \gamma_{21})$ using a reduced two-mode autoparametric resonance model. We show that sensitivity organizes into well-defined geometric structures in parameter space that cannot be inferred from comb existence conditions alone. Primary detuning $\Delta_1$ governs smooth, spatially extended responses that preserve dynamical continuity, while secondary detuning $\Delta_2$ produces sharply localized sensitivity along resonance manifolds tied to the onset and saturation of autoparametric excitation. The drive amplitude $f$ sets both the activation threshold and the global amplification scale, with peak sensitivity emerging near threshold rather than deep in the nonlinear regime. The relative damping $\gamma_{21}$, by contrast, continuously redistributes energy between modes without introducing discontinuities. Taken together, these results show that PFC systems actively organize their own sensitivity through intrinsic nonlinear dynamics, offering a principled foundation for sensitivity engineering in next-generation phononic sensing platforms.

\section{Mathematical Model}

The equations of motion corresponding to two coupled modes are given by
\begin{align}
\ddot{x}_1 + 2\gamma_1 \dot{x}_1 + \omega_1^2 x_1 + \alpha_{22} x_2^2 &= F \cos(\omega_D t) \tag{1} \\
\ddot{x}_2 + 2\gamma_2 \dot{x}_2 + \omega_2^2 x_2 + \alpha_{12} x_1 x_2 &= 0 \tag{2}
\end{align}
where $\gamma_1$ and $\gamma_2$ are damping coefficients, $\omega_1$ and $\omega_2$ are resonance frequencies, $\alpha_{12}$ and $\alpha_{22}$ are nonlinear modal coupling coefficients, and $F$ and $\omega_D$ are the amplitude and frequency of the drive respectively \cite{b28}. The coupling structure in (1)-(2) reflects a quadratic nonlinearity arising from the nonlinear strain relationship between the two modes \cite{b28}, and the Duffing nonlinearity is omitted since PFCs are observed experimentally before third-order effects become significant \cite{b3}.

The modes are written as
\begin{align}
x_1 &= \tfrac{1}{2}\big(u_1 e^{i\omega_D t} + u_1^{*} e^{-i\omega_D t}\big) \tag{3} \\
x_2 &= \tfrac{1}{2}\big(u_2 e^{i\omega_D t/2} + u_2^{*} e^{-i\omega_D t/2}\big) \tag{4}
\end{align}
This ansatz encodes the 2:1 autoparametric resonance condition $\omega_1 \approx 2\omega_2$:
the nonlinear coupling in (2) can parametrically excite mode 2 at the subharmonic
$\omega_D/2$, and it is precisely this subharmonic response that seeds the comb sidebands
observed in the spectra. The complex envelopes $u_1$ and $u_2$ are assumed to vary slowly
relative to the carrier comb lines, that is
\begin{equation}
|\ddot{u}_j| \ll \omega_D|\dot{u}_j|, \qquad
|\dot{u}_j| \ll \omega_j|u_j|, \qquad
\gamma_j \ll \omega_j, \quad j = 1,2. \tag{$*$}
\end{equation}
Under these conditions, second-order time derivatives of $u_j$ and products of damping
with first-order derivatives are negligible compared to the dominant carrier-frequency
terms.

We now collect the coefficients of $e^{i\omega_D t}$ in (1) and those of
$e^{i\omega_D t/2}$ in (2) to obtain the following relationships:
\begin{equation}
\begin{aligned}
\ddot{u}_1
&+ \big[ 2\gamma_1 + 2 i \omega_D \big] \dot{u}_1
+ \big[ (\omega_1^2 - \omega_D^2) + 2 i \gamma_1 \omega_D \big] u_1 \\
&+ \frac{\alpha_{22}}{2} u_2^2
= F
\end{aligned}
\tag{5}
\end{equation}
\begin{equation}
\begin{aligned}
\ddot{u}_2
&+ \big[ 2\gamma_2 + i \omega_D \big] \dot{u}_2
+ \big[ (\omega_2^2 - \omega_D^2/4) + i \gamma_2 \omega_D \big] u_2 \\
&+ \frac{\alpha_{12}}{2} u_1 u_2^{*}
= 0
\end{aligned}
\tag{6}
\end{equation}
Applying the slowly varying envelope conditions $(*)$ to (5)-(6) and driving near the
resonance of mode 1 (i.e., $\omega_D \approx \omega_1$), the second-order derivative
terms drop out and the equations reduce to first order in the slow time. We then apply
the following normalizations:
\begin{align*}
\tau &= \gamma_1 t, \quad
f = \frac{\alpha_{12}}{8 \omega_D^2 \gamma_1^2} F, \quad
\gamma_{21} = \frac{\gamma_2}{\gamma_1} \\
\Delta_1 &= \frac{\omega_D - \omega_1}{\gamma_1}, \quad
\Delta_2 = \frac{\omega_D - 2\omega_2}{2\gamma_1} = \frac{\Delta_1}{2} + \kappa \\
\psi_1 &= \frac{\alpha_{12}}{4 \gamma_1 \omega_D} u_1, \quad
\psi_2 = \frac{\sqrt{\alpha_{12} \alpha_{22}}}{4 \gamma_2 \omega_D} u_2
\end{align*}
where $\kappa = (\omega_1 - 2\omega_2)/(2\gamma_1)$ is the normalized frequency mismatch
from perfect 2:1 resonance: $\kappa = 0$ corresponds to exact internal resonance, and
the sign of $\kappa$ determines whether the comb existence region falls on the red- or
blue-detuned side of mode 1, as established in \cite{b28}. The normalized drive $f$
scales the external forcing by the nonlinear coupling strength and damping, so variations
in $f$ in the results below correspond physically to changes in the drive force. This
yields the reduced system
\begin{align}
\frac{\partial \psi_1}{\partial \tau} &= -i f - \big(1 + i \Delta_1\big)\psi_1
+ i \psi_2^2 \tag{7} \\
\frac{\partial \psi_2}{\partial \tau} &= -\big(\gamma_{21} + i \Delta_2\big)\psi_2
+ 2 i \psi_1 \psi_2^{*} \tag{8}
\end{align}

The solutions of equations (7) and (8) take the form of frequency combs as shown in \cite{b28}. This paper numerically investigates the sensitivity of these frequency combs to small perturbations in the system parameters viz. $\Delta_1$, $\Delta_2$ and $\gamma_{21}$.

\section{Results and Discussion}

\begin{figure*}
\centering
\includegraphics[width=0.7\textwidth]{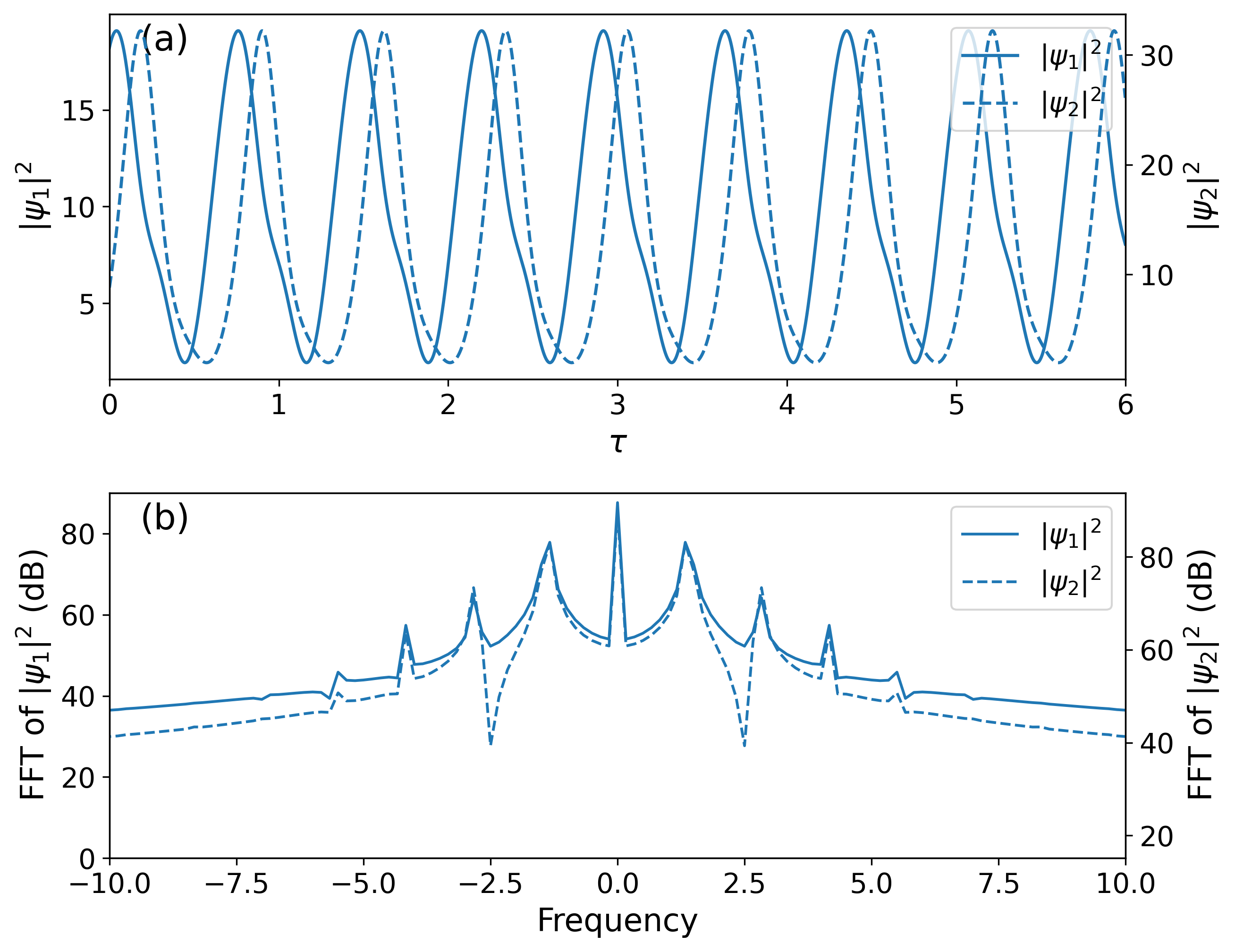}
\caption{(a) Time-domain evolution of the coupled mode intensities $|\psi_1|^2$ and $|\psi_2|^2$ plotted together; (b) Corresponding Fourier spectra (FFT) of the same quantities. Here, $\tau$ and frequency are unitless normalized variables. The simulation parameters are $\Delta_1 = 5$, $\Delta_2 = -6.5$, $\gamma_{21} = 1$, and $f = 24$ with initial conditions are $\psi_1(0) = 5$ and $\psi_2(0) = 3$.}
\label{fig1}
\end{figure*}

\begin{figure*}
\centering
\includegraphics[width=0.9\textwidth]{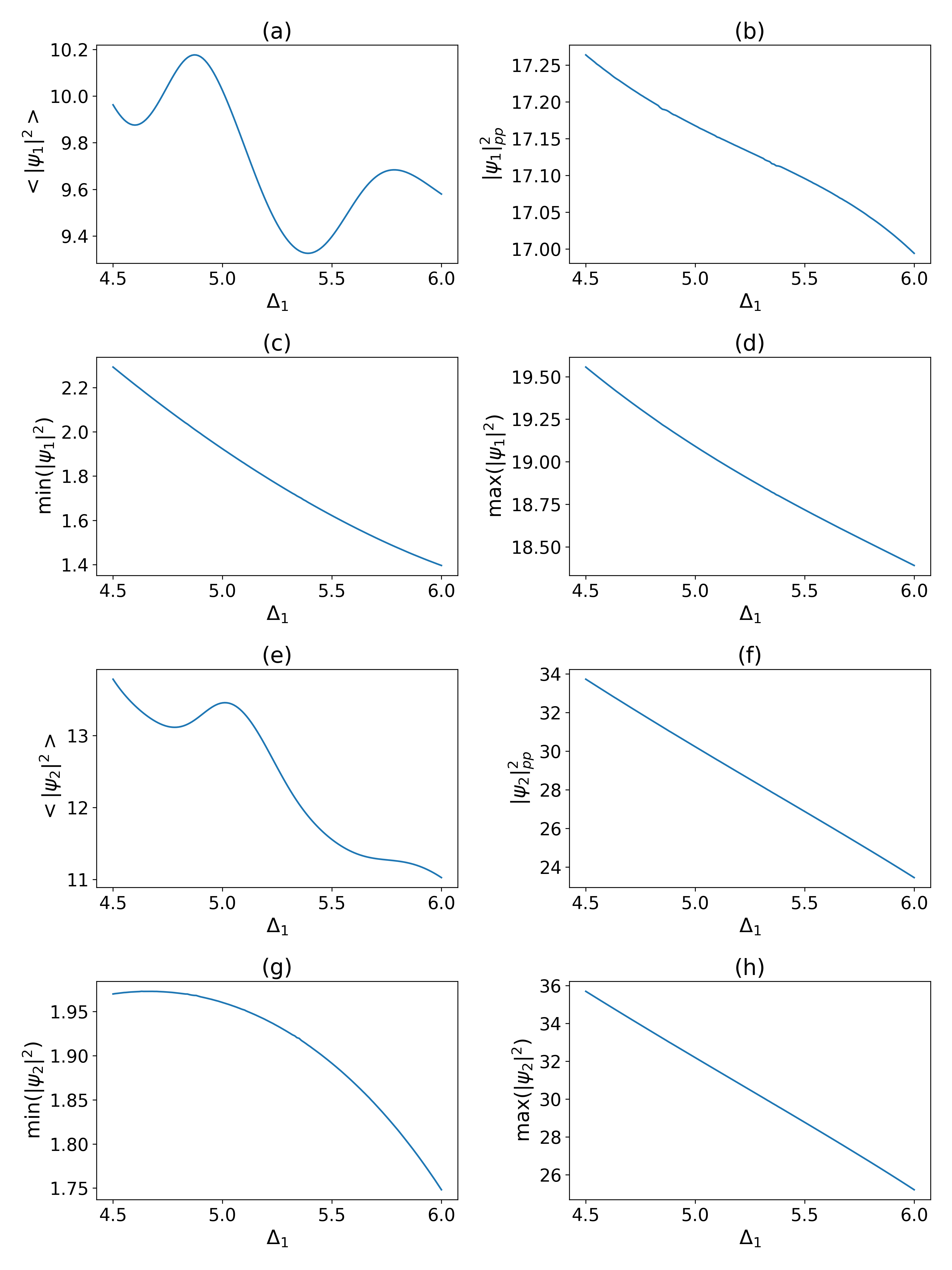}
\caption{Evolution of modal intensities of $\psi_1$ and $\psi_2$ as a function of detuning $\Delta_{1}$, with $\Delta_{2} = -6.5$, $\gamma_{21} = 1$, and driving amplitude $f = 24$ held fixed. Panels (a)-(d) correspond to mode $\psi_{1}$, showing the mean intensity $\langle |\psi_1|^2 \rangle$, peak-to-peak intensity $|\psi_1|^2_{\mathrm{pp}}$, minimum intensity $\min(|\psi_1|^2)$, and maximum intensity $\max(|\psi_1|^2)$, respectively. Panels (e)-(h) show the corresponding quantities for mode $\psi_{2}$.}
\label{fig2}
\end{figure*}

\begin{figure*}
\centering
\includegraphics[width=0.9\textwidth]{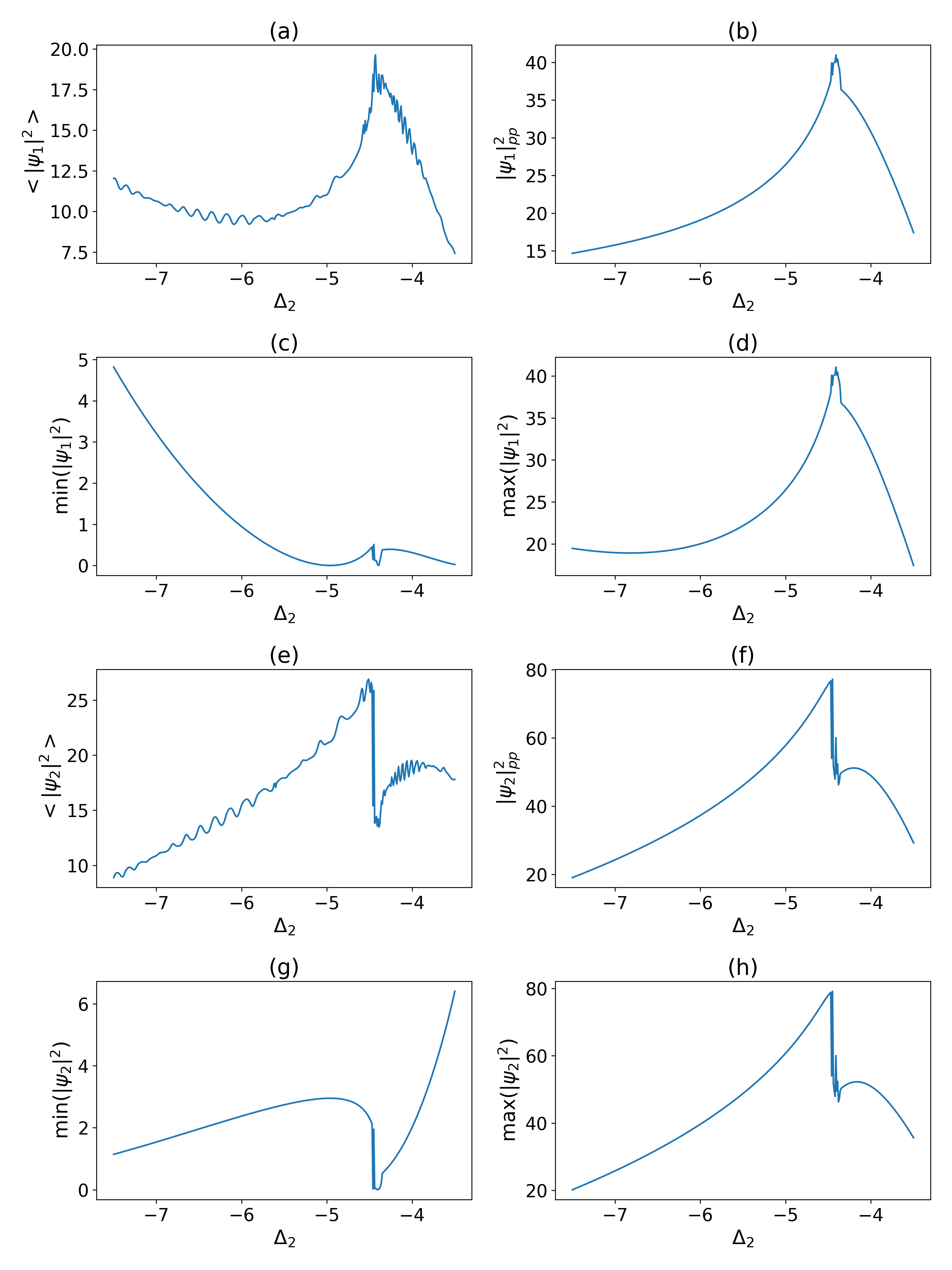}
\caption{Evolution of modal intensities of $\psi_1$ and $\psi_2$ as a function of detuning $\Delta_{2}$, with $\Delta_{1} = 5$, $\gamma_{21} = 1$, and driving amplitude $f = 24$ held fixed. Panels (a)-(d) correspond to mode $\psi_{1}$, showing the mean intensity $\langle |\psi_1|^2 \rangle$, peak-to-peak intensity $|\psi_1|^2_{\mathrm{pp}}$, minimum intensity $\min(|\psi_1|^2)$, and maximum intensity $\max(|\psi_1|^2)$, respectively. Panels (e)-(h) show the corresponding quantities for mode $\psi_{2}$.}
\label{fig3}
\end{figure*}

\begin{figure*}
\centering
\includegraphics[width=0.9\textwidth]{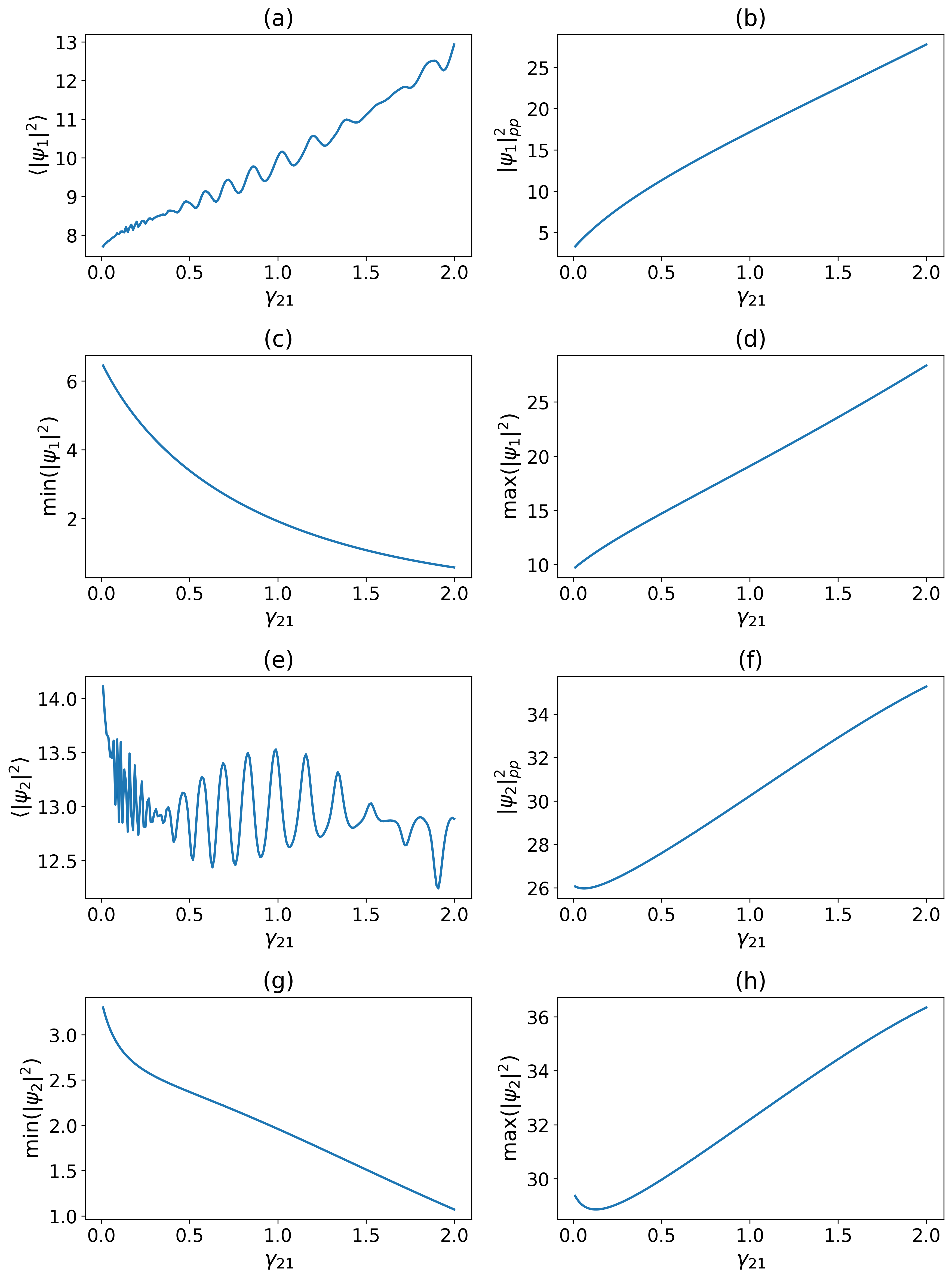}
\caption{Evolution of modal intensities of $\psi_1$ and $\psi_2$ as a function of the relative damping $\gamma_{21}$, with $\Delta_{1} = 5$, $\Delta_{2} = -6.5$, and driving amplitude $f = 24$ held fixed. Panels (a)-(d) correspond to mode $\psi_{1}$, showing the mean intensity $\langle |\psi_1|^2 \rangle$, peak-to-peak intensity $|\psi_1|^2_{\mathrm{pp}}$, minimum intensity $\min(|\psi_1|^2)$, and maximum intensity $\max(|\psi_1|^2)$, respectively. Panels (e)-(h) show the corresponding quantities for mode $\psi_{2}$.}
\label{fig4}
\end{figure*}

\begin{figure*}
    \centering
    \includegraphics[width=0.95\linewidth]{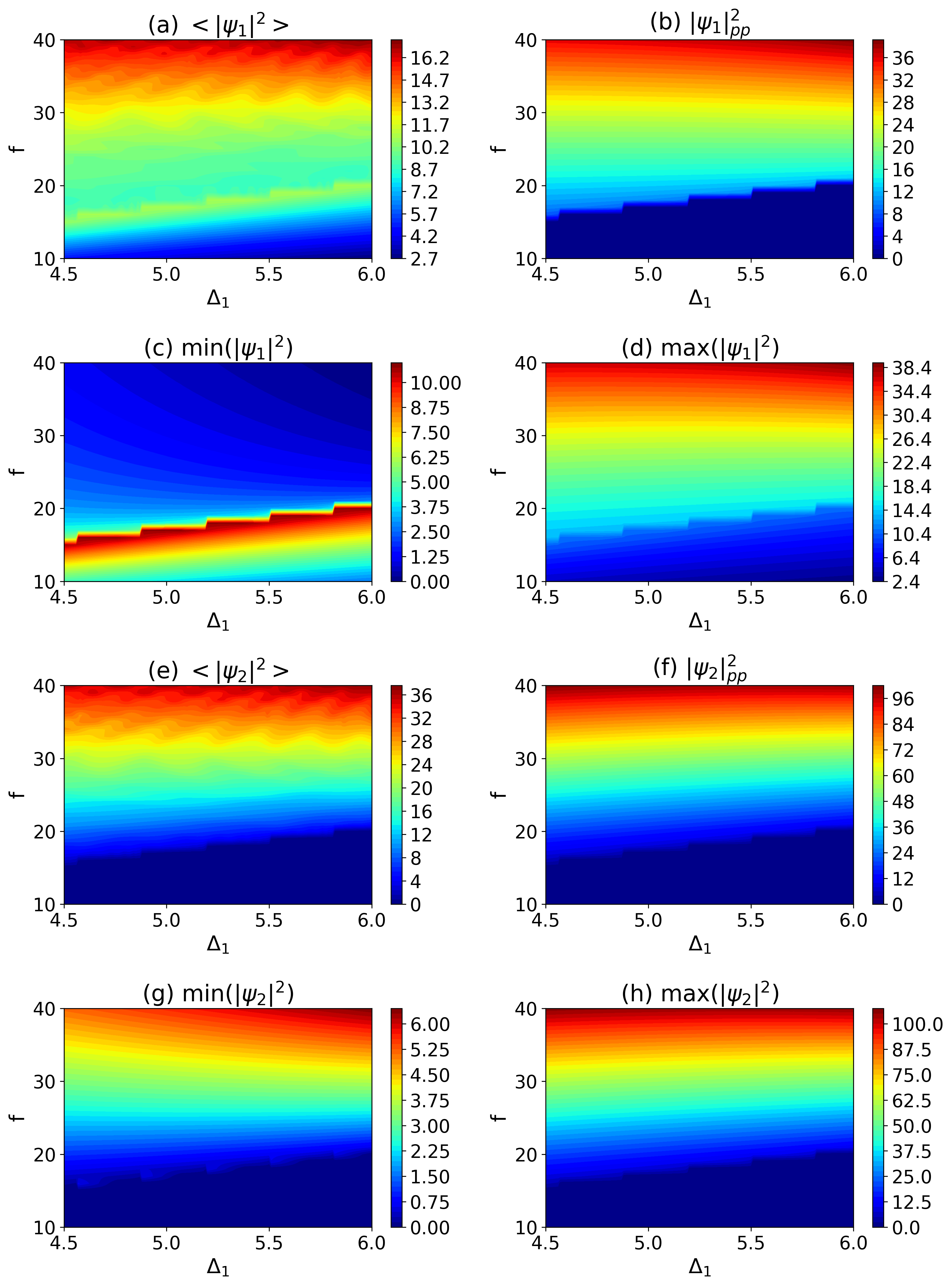}
    \caption{Evolution of modal intensities as a function of detuning $\Delta_{1}$ and driving amplitude $f$ for the two-mode system, with $\Delta_{2} = -6.5$ and $\gamma_{21} = 1$ held fixed. Panels (a)-(d) correspond to mode $\psi_{1}$, showing the mean intensity $\langle |\psi_1|^2 \rangle$, peak-to-peak intensity $|\psi_1|^2_{\mathrm{pp}}$, minimum intensity $\min(|\psi_1|^2)$, and maximum intensity $\max(|\psi_1|^2)$, respectively. Panels (e)-(h) show the corresponding quantities for mode $\psi_{2}$.}
    \label{fig5}
\end{figure*}

\begin{figure*}
    \centering
    \includegraphics[width=0.95\linewidth]{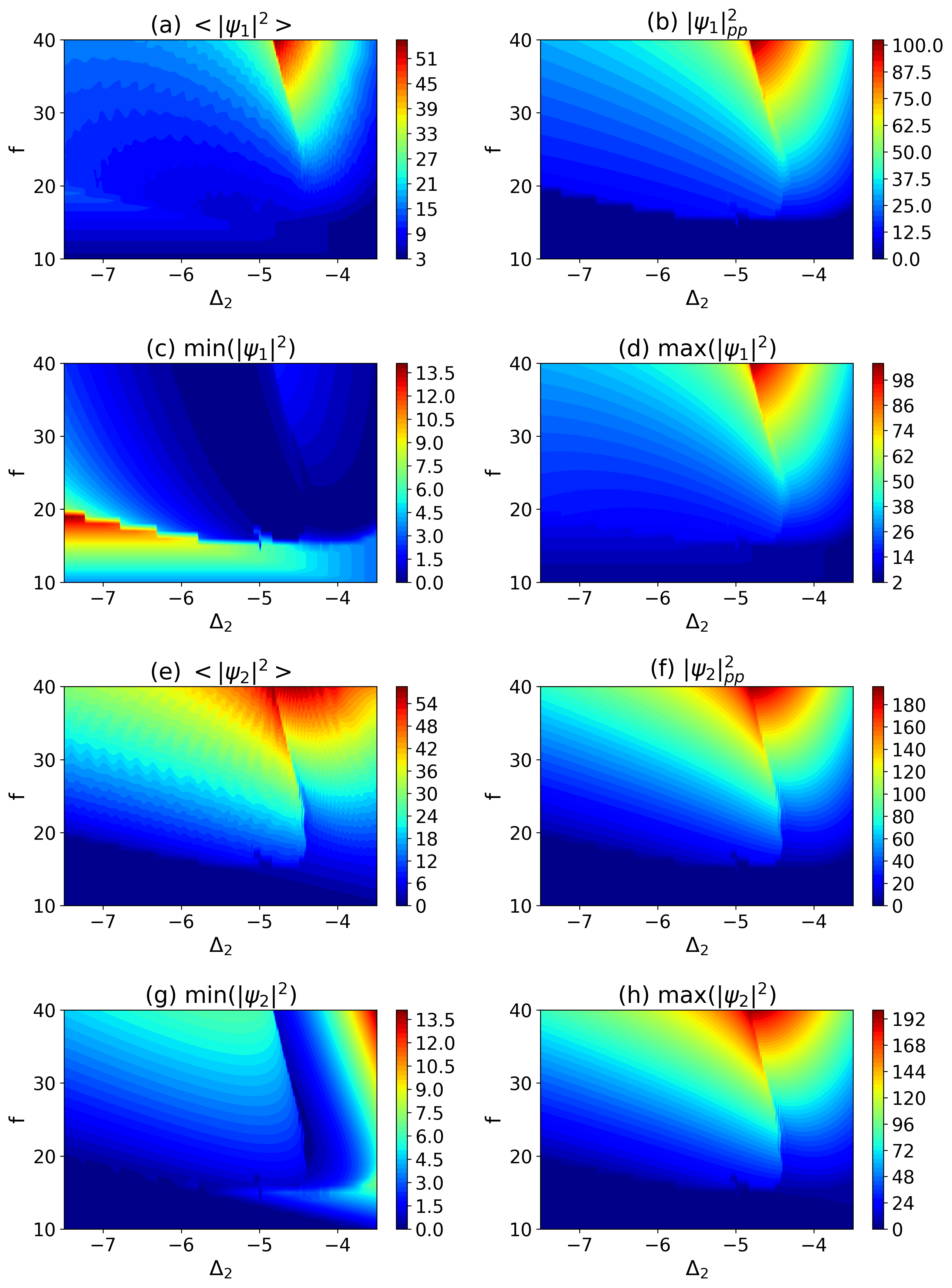}
    \caption{Evolution of modal intensities as a function of detuning $\Delta_{2}$ and driving amplitude $f$ for the two-mode system, with $\Delta_{1} = 5$ and $\gamma_{21} = 1$ held fixed. Panels (a)-(d) correspond to mode $\psi_{1}$, showing the mean intensity $\langle |\psi_1|^2 \rangle$, peak-to-peak intensity $|\psi_1|^2_{\mathrm{pp}}$, minimum intensity $\min(|\psi_1|^2)$, and maximum intensity $\max(|\psi_1|^2)$, respectively. Panels (e)-(h) show the corresponding quantities for mode $\psi_{2}$.}
    \label{fig6}
\end{figure*}

\begin{figure*}
\centering
\includegraphics[width=0.95\textwidth]{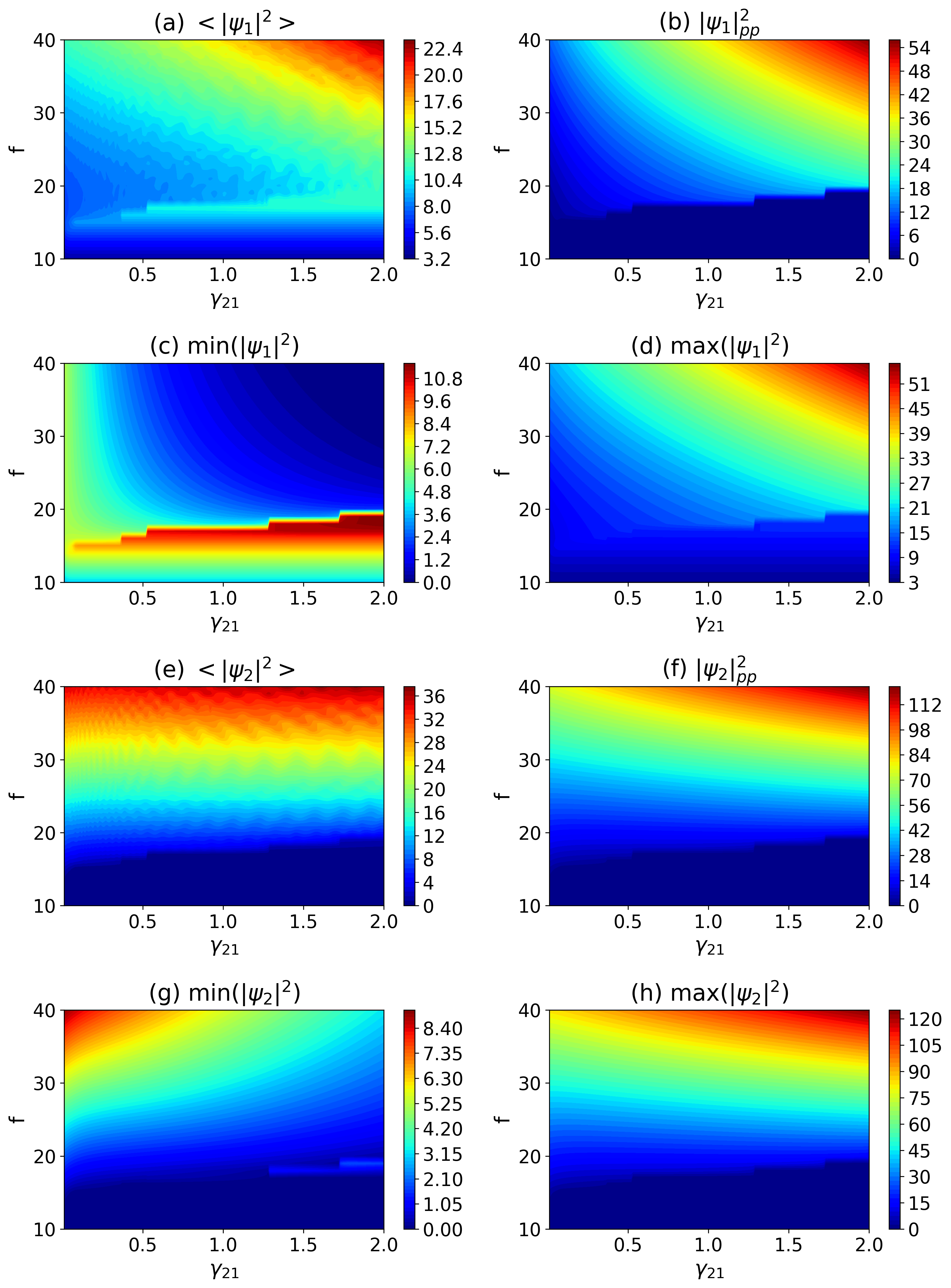}
\caption{Evolution of modal intensities as a function of the relative damping $\gamma_{21}$ and driving amplitude $f$, with $\Delta_{1} = 5$ and $\Delta_{2} = -6.5$ held fixed. Panels (a)-(d) correspond to mode $\psi_{1}$, showing the mean intensity $\langle |\psi_1|^2 \rangle$, peak-to-peak intensity $|\psi_1|^2_{\mathrm{pp}}$, minimum intensity $\min(|\psi_1|^2)$, and maximum intensity $\max(|\psi_1|^2)$, respectively. Panels (e)-(h) show the corresponding quantities for mode $\psi_{2}$.}
\label{fig7}
\end{figure*}

\begin{figure*}
    \centering
    \includegraphics[width=0.95\linewidth]{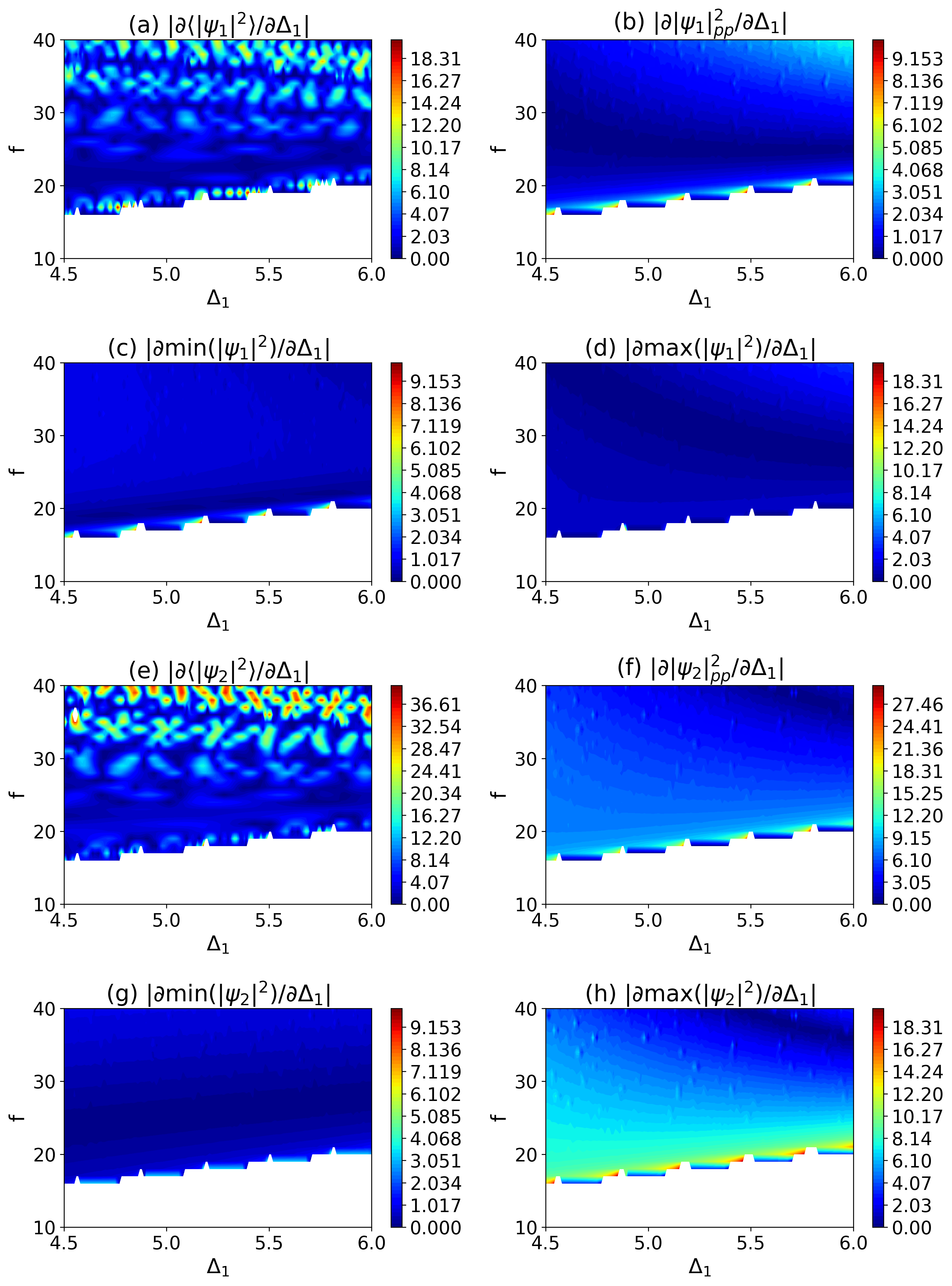}
    \caption{Absolute differential sensitivity of modal intensities to variations in detuning $\Delta$ (with $\Delta \equiv \Delta_{1}$) as a function of driving amplitude $f$, with $\Delta_{2} = -6.5$ and $\gamma_{21} = 1$ held fixed. Panels (a)-(d) correspond to mode $\psi_{1}$, showing $|\partial \langle |\psi_1|^2 \rangle / \partial \Delta|$, $|\partial |\psi_1|^2_{\mathrm{pp}} / \partial \Delta|$, $|\partial \min(|\psi_1|^2) / \partial \Delta|$, and $|\partial \max(|\psi_1|^2) / \partial \Delta|$, respectively. Panels (e)-(h) show the corresponding quantities for mode $\psi_{2}$.}
    \label{fig8}
\end{figure*}

\begin{figure*}
    \centering
    \includegraphics[width=0.95\linewidth]{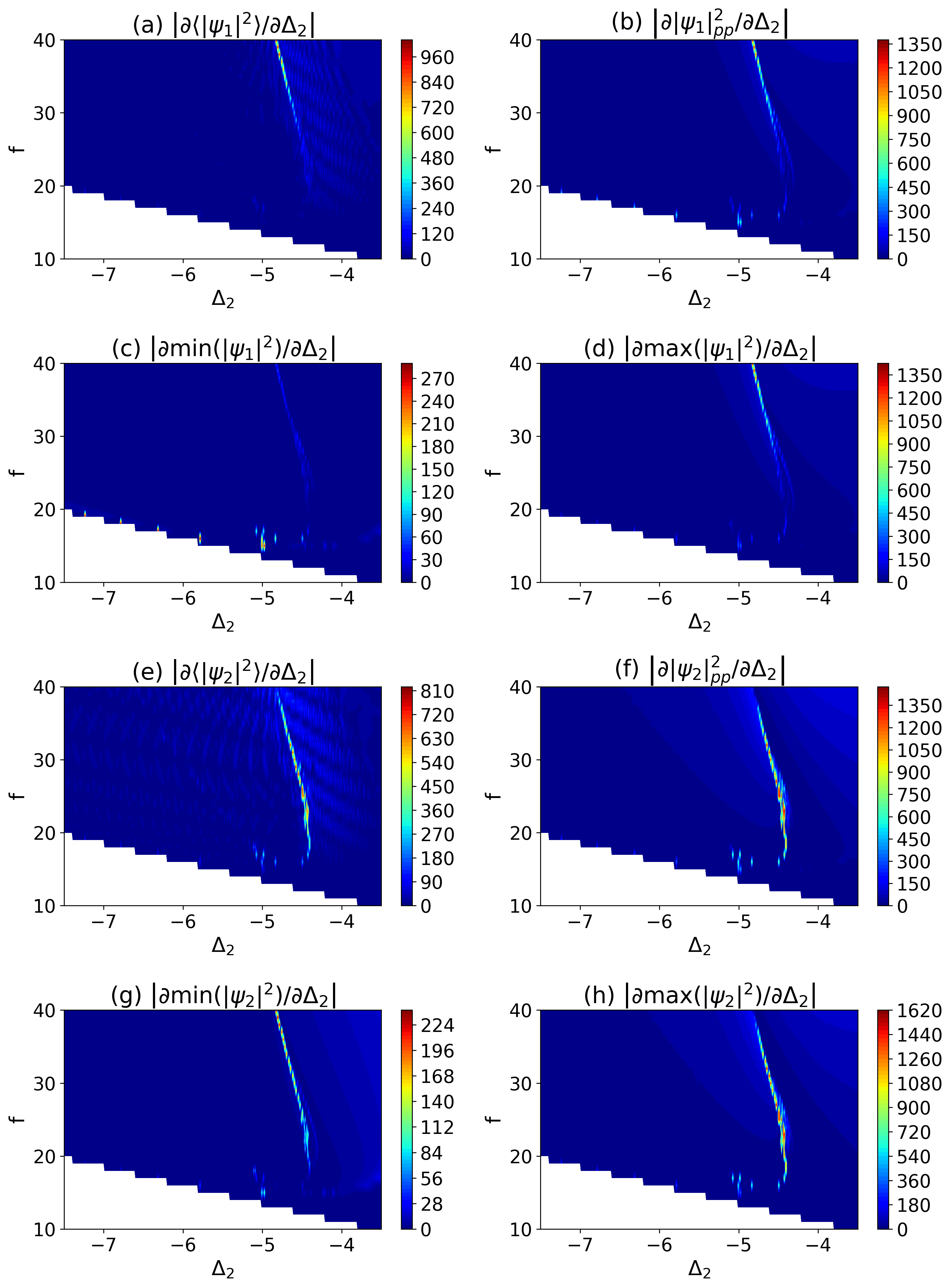}
    \caption{Absolute differential sensitivity of modal intensities to variations in detuning $\Delta$ (with $\Delta \equiv \Delta_{2}$) as a function of driving amplitude $f$, with $\Delta_{1} = 5$ and $\gamma_{21} = 1$ held fixed. Panels (a)-(d) correspond to mode $\psi_{1}$, showing $|\partial \langle |\psi_1|^2 \rangle / \partial \Delta|$, $|\partial |\psi_1|^2_{\mathrm{pp}} / \partial \Delta|$, $|\partial \min(|\psi_1|^2) / \partial \Delta|$, and $|\partial \max(|\psi_1|^2) / \partial \Delta|$, respectively. Panels (e)-(h) show the corresponding quantities for mode $\psi_{2}$.}
    \label{fig9}
\end{figure*}

\begin{figure*}
\centering
\includegraphics[width=0.95\textwidth]{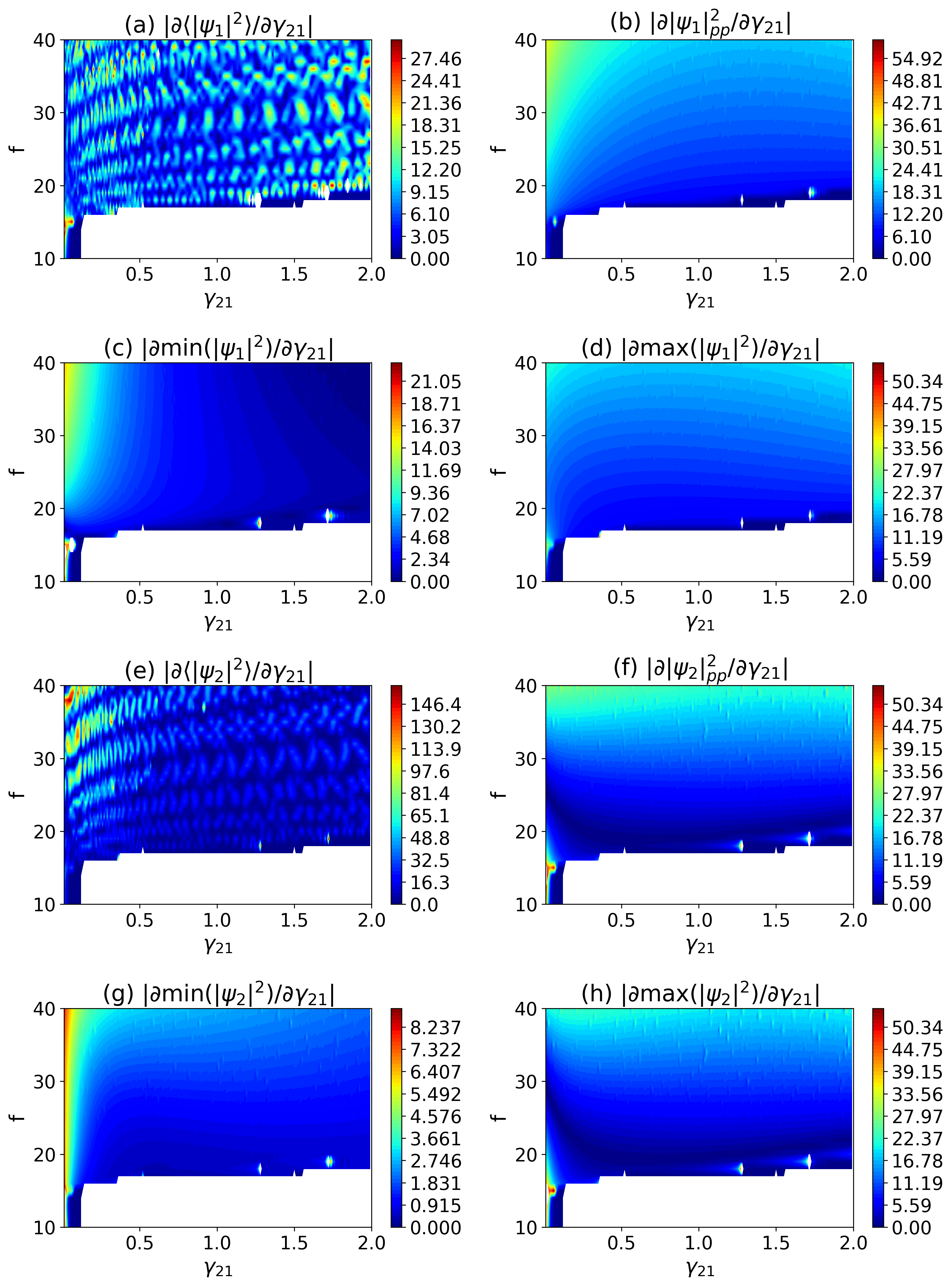}
\caption{Absolute differential sensitivity of modal intensities with respect to variations in the relative damping $\gamma_{21}$ as a function of driving amplitude $f$, with $\Delta_{1} = 5$ and $\Delta_{2} = -6.5$ held fixed. Panels (a)-(d) correspond to mode $\psi_{1}$, showing $|\partial \langle |\psi_1|^2 \rangle / \partial \gamma_{21}|$, $|\partial |\psi_1|^2_{\mathrm{pp}} / \partial \gamma_{21}|$, $|\partial \min(|\psi_1|^2) / \partial \gamma_{21}|$, and $|\partial \max(|\psi_1|^2) / \partial \gamma_{21}|$, respectively. Panels (e)-(h) show the corresponding quantities for mode $\psi_{2}$.}
\label{fig10}
\end{figure*}

\begin{figure*}
    \centering
    \includegraphics[width=0.95\linewidth]{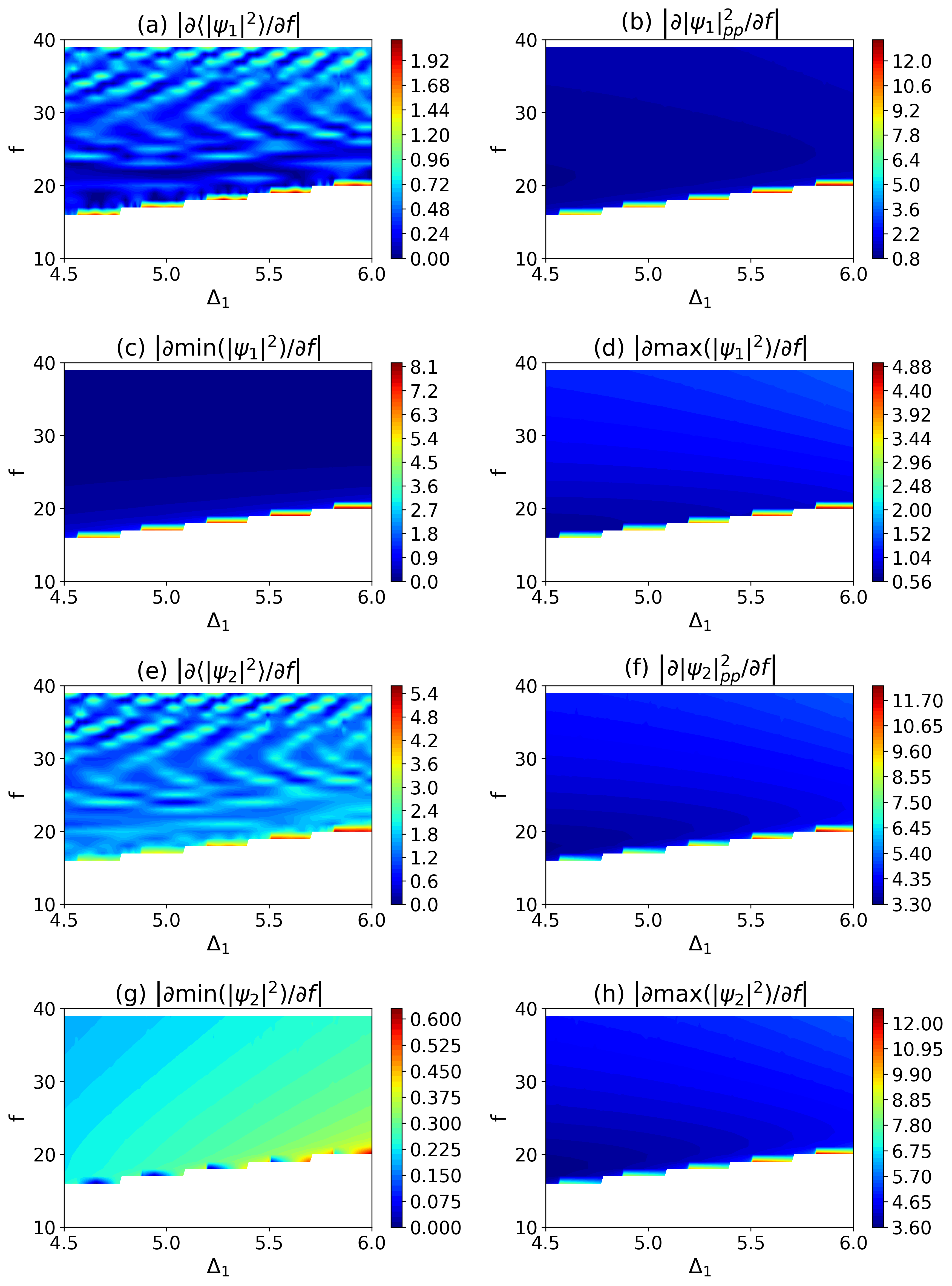}
    \caption{Absolute differential sensitivity of modal intensities to variations in driving amplitude $f$ as a function of detuning $\Delta_{1}$, with $\Delta_{2} = -6.5$ and $\gamma_{21} = 1$ held fixed. Panels (a)-(d) correspond to mode $\psi_{1}$, showing $|\partial \langle |\psi_1|^2 \rangle / \partial f|$, $|\partial |\psi_1|^2_{\mathrm{pp}} / \partial f|$, $|\partial \min(|\psi_1|^2) / \partial f|$, and $|\partial \max(|\psi_1|^2) / \partial f|$, respectively. Panels (e)-(h) show the corresponding quantities for mode $\psi_{2}$.}
    \label{fig11}
\end{figure*}

\begin{figure*}
    \centering
    \includegraphics[width=0.95\linewidth]{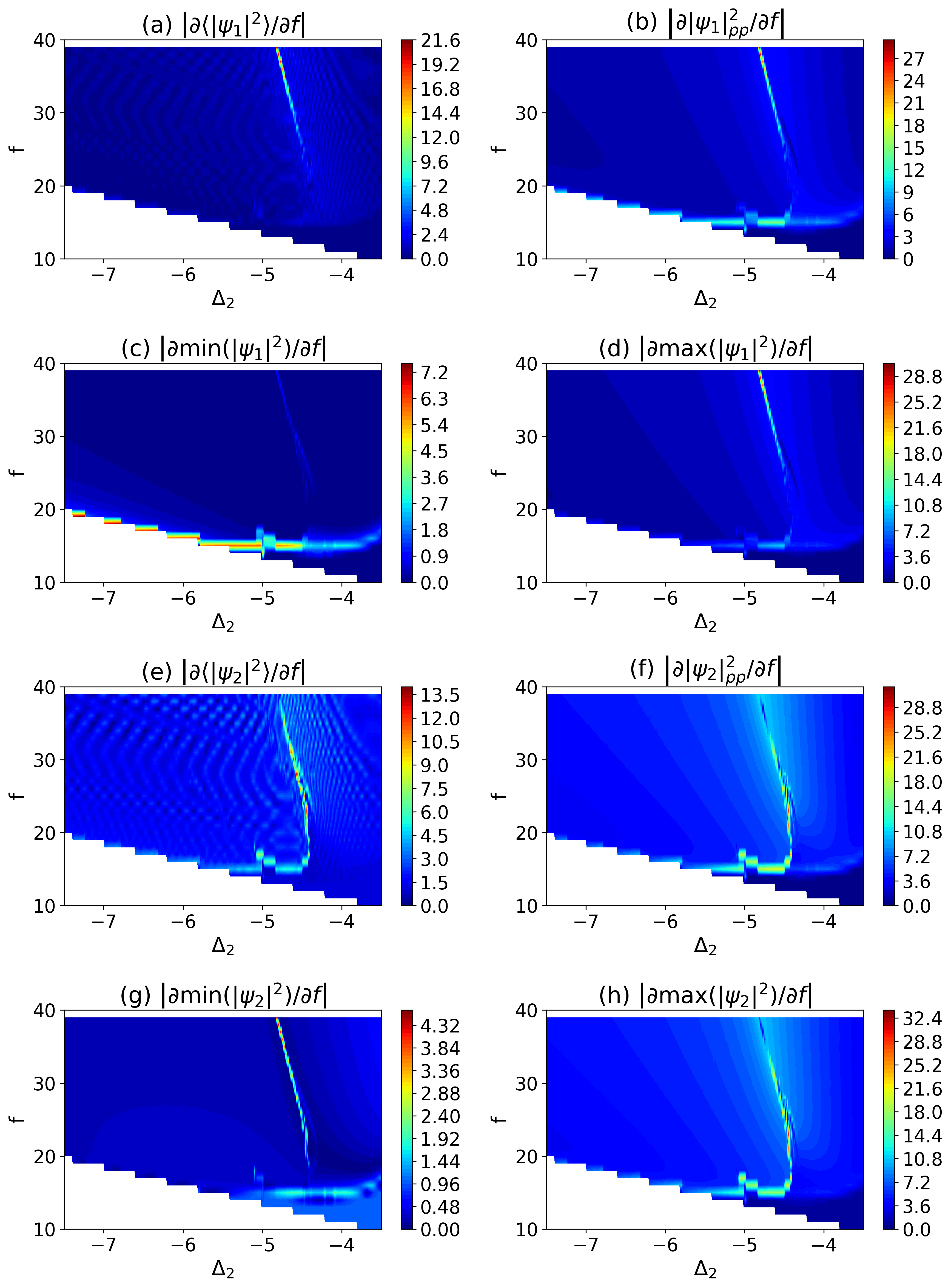}
    \caption{Absolute differential sensitivity of modal intensities to variations in driving amplitude $f$ as a function of detuning $\Delta_{2}$, with $\Delta_{1} = 5$ and $\gamma_{21} = 1$ held fixed. Panels (a)-(d) correspond to mode $\psi_{1}$, showing $|\partial \langle |\psi_1|^2 \rangle / \partial f|$, $|\partial |\psi_1|^2_{\mathrm{pp}} / \partial f|$, $|\partial \min(|\psi_1|^2) / \partial f|$, and $|\partial \max(|\psi_1|^2) / \partial f|$, respectively. Panels (e)-(h) show the corresponding quantities for mode $\psi_{2}$.}
    \label{fig12}
\end{figure*}

\begin{figure*}
\centering
\includegraphics[width=0.95\textwidth]{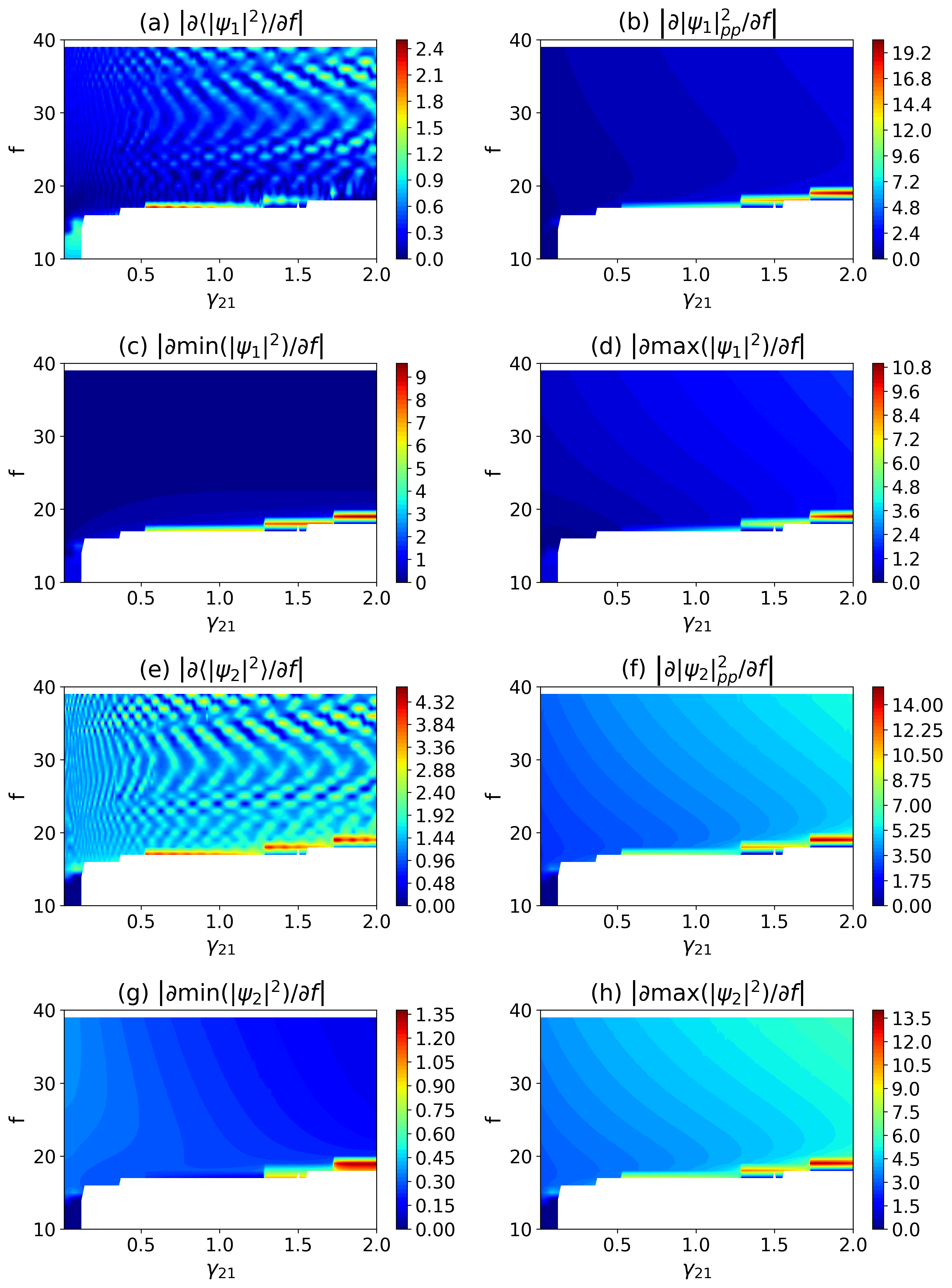}
\caption{Absolute differential sensitivity of modal intensities to variations in the driving amplitude $f$ as a function of the relative damping $\gamma_{21}$, with $\Delta_{1} = 5$ and $\Delta_{2} = -6.5$ held fixed. Panels (a)-(d) correspond to mode $\psi_{1}$, showing $|\partial \langle |\psi_1|^2 \rangle / \partial f|$, $|\partial |\psi_1|^2_{\mathrm{pp}} / \partial f|$, $|\partial \min(|\psi_1|^2) / \partial f|$, and $|\partial \max(|\psi_1|^2) / \partial f|$, respectively. Panels (e)-(h) show the corresponding quantities for mode $\psi_{2}$.}
\label{fig13}
\end{figure*}

\begin{figure*}
\centering
\includegraphics[width=0.75\textwidth]{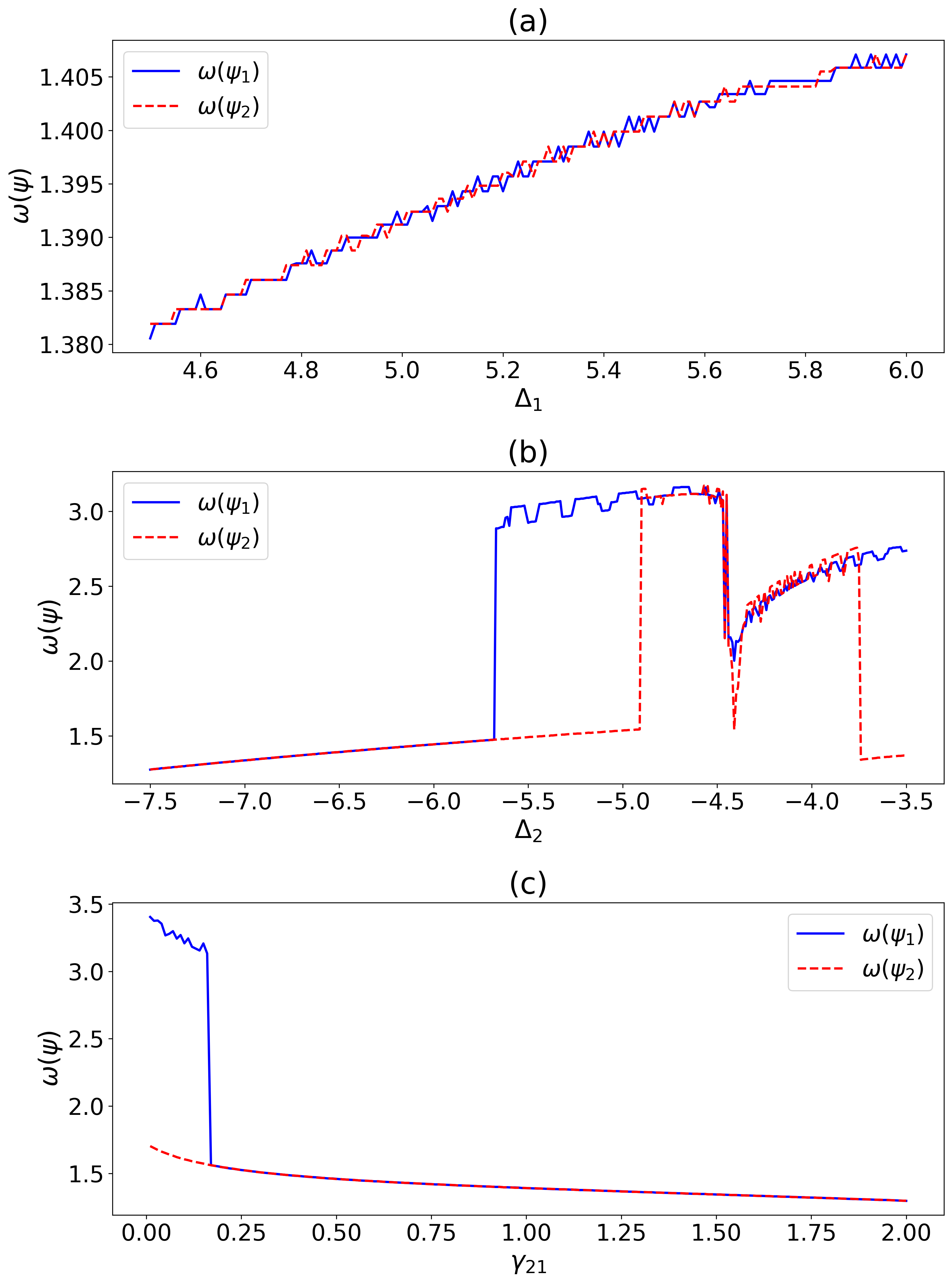}
\caption{Evolution of the frequency $\omega(\psi)$ as a function of detuning and relative damping at fixed drive amplitude $f=24$ with the solid blue curve represents $\omega(\psi_{1})$, while the dashed red curve denotes $\omega(\psi_{2})$. (a) Frequency variation with $\Delta_{1}$ at fixed $\Delta_{2} = -6.5$ and $\gamma_{21} = 1$. (b) Frequency variation with $\Delta_{2}$ at fixed $\Delta_{1} = 5$ and $\gamma_{21} = 1$. (c) Frequency variation with relative damping $\gamma_{21}$ at fixed $\Delta_{1} = 5$ and $\Delta_{2} = -6.5$.}
\label{fig14}
\end{figure*}

\begin{figure*}
\centering
\includegraphics[width=0.95\textwidth]{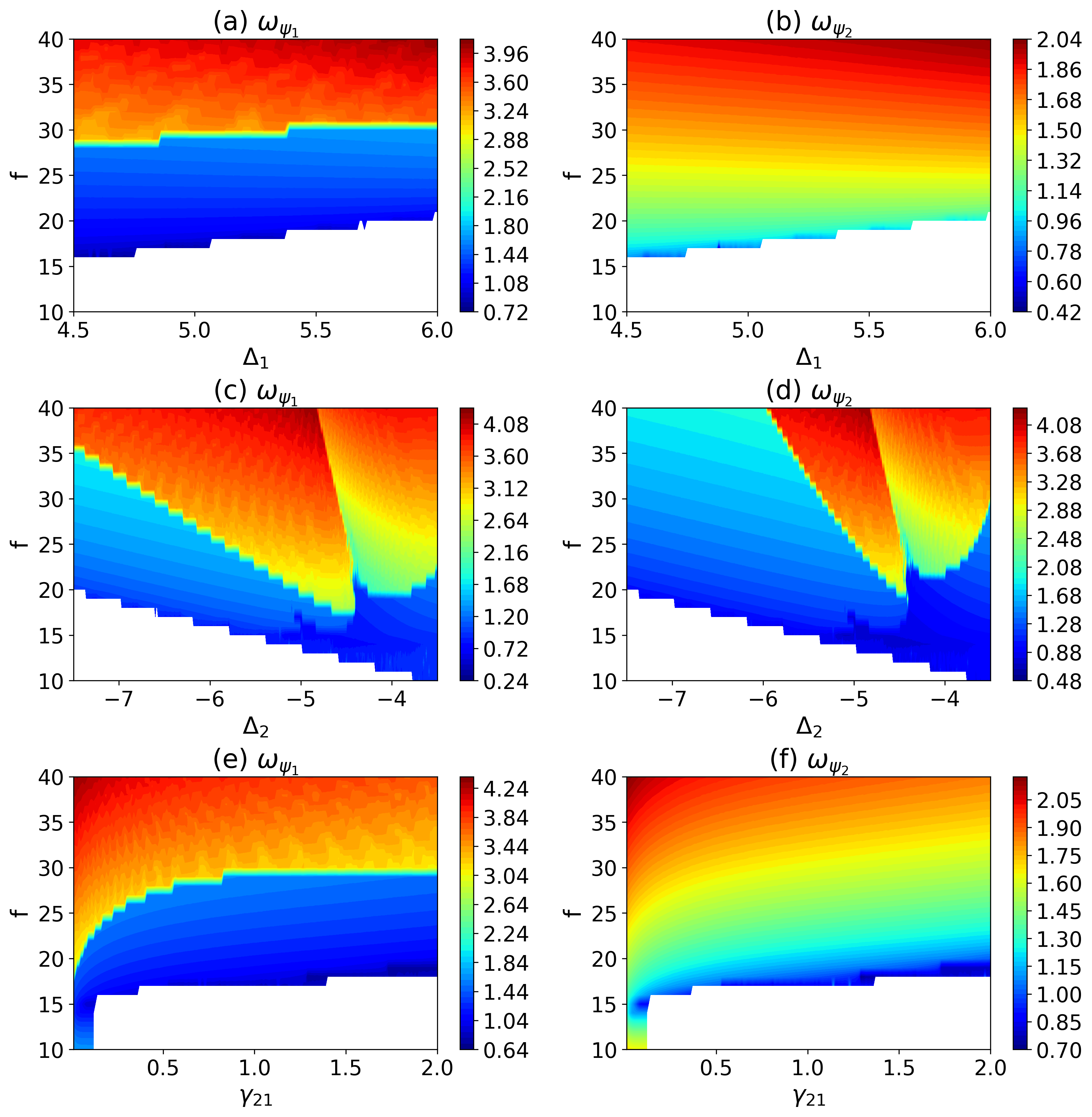}
\caption{Oscillation frequencies of the comb modes as functions of detuning parameters and driving amplitude $f$. (a),(b) $\omega_{\psi_1}$ and $\omega_{\psi_2}$ as functions of $\Delta_1$ and $f$, with $\Delta_2 = -6.5$ and $\gamma_{21} = 1$ held fixed. (c),(d) $\omega_{\psi_1}$ and $\omega_{\psi_2}$ as functions of $\Delta_2$ and $f$, with $\Delta_1 = 5$ and $\gamma_{21} = 1$ held fixed. (e),(f) $\omega_{\psi_1}$ and $\omega_{\psi_2}$ as functions of the relative damping $\gamma_{21}$ and $f$, with $\Delta_1 = 5$ and $\Delta_2 = -6.5$ held fixed.}
\label{fig15}
\end{figure*}

\begin{figure*}
\centering
\includegraphics[width=0.95\textwidth]{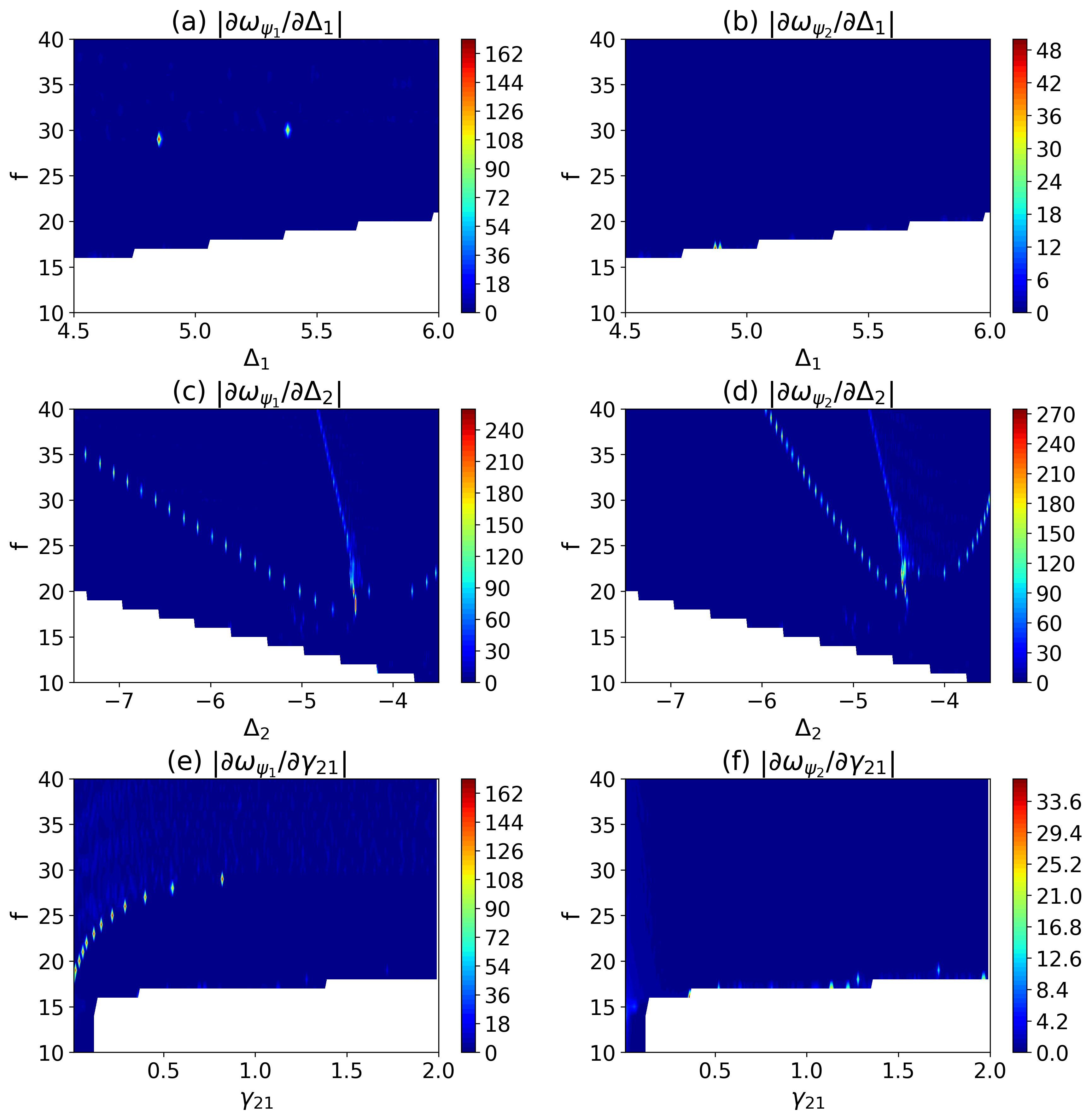}
\caption{Absolute differential sensitivity of the comb mode frequencies with respect to system parameters as functions of the driving amplitude $f$. (a),(b) $|\partial \omega_{\psi_1}/\partial \Delta_1|$ and $|\partial \omega_{\psi_2}/\partial \Delta_1|$, with $\Delta_2 = -6.5$ and $\gamma_{21} = 1$ held fixed. (c),(d) $|\partial \omega_{\psi_1}/\partial \Delta_2|$ and $|\partial \omega_{\psi_2}/\partial \Delta_2|$, with $\Delta_1 = 5$ and $\gamma_{21} = 1$ held fixed. (e),(f) $|\partial \omega_{\psi_1}/\partial \gamma_{21}|$ and $|\partial \omega_{\psi_2}/\partial \gamma_{21}|$, with $\Delta_1 = 5$ and $\Delta_2 = -6.5$ held fixed.}
\label{fig16}
\end{figure*}

\begin{figure*}
\centering
\includegraphics[width=0.95\textwidth]{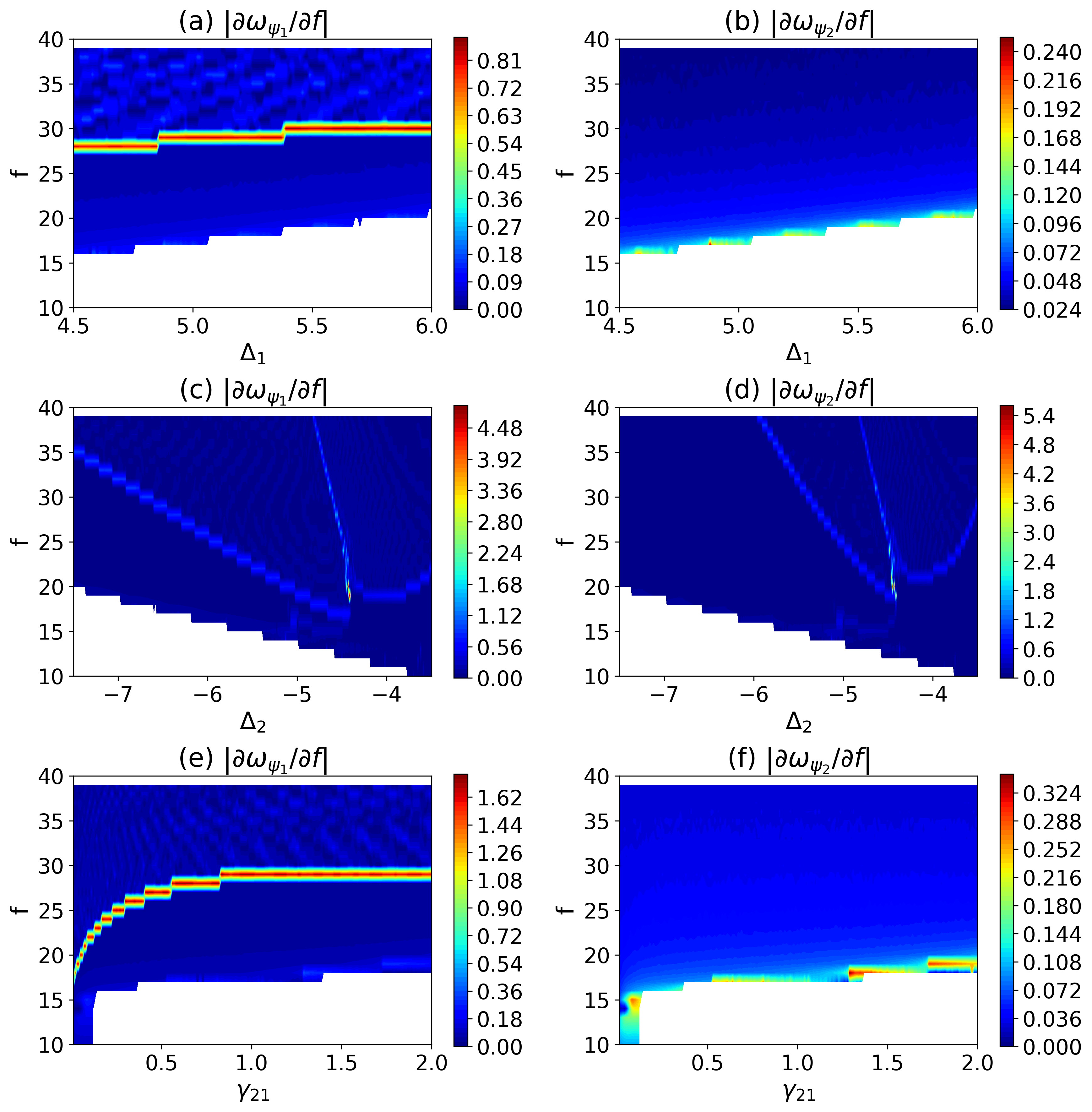}
\caption{Absolute differential sensitivity of the comb mode frequencies to variations in the driving amplitude $f$. (a),(b) $|\partial \omega_{\psi_1}/\partial f|$ and $|\partial \omega_{\psi_2}/\partial f|$ as functions of $\Delta_1$ and $f$, with $\Delta_2 = -6.5$ and $\gamma_{21} = 1$ held fixed. (c),(d) $|\partial \omega_{\psi_1}/\partial f|$ and $|\partial \omega_{\psi_2}/\partial f|$ as functions of $\Delta_2$ and $f$, with $\Delta_1 = 5$ and $\gamma_{21} = 1$ held fixed. (e),(f) $|\partial \omega_{\psi_1}/\partial f|$ and $|\partial \omega_{\psi_2}/\partial f|$ as functions of $\gamma_{21}$ and $f$, with $\Delta_1 = 5$ and $\Delta_2 = -6.5$ held fixed.}
\label{fig17}
\end{figure*}

Figure~\ref{fig1} demonstrates the steady-state dynamics of the coupled modes in the time and frequency domains. Panel (a) plots the squared amplitudes $|\psi_1|^2$ and $|\psi_2|^2$ against dimensionless time $\tau$. Both modes oscillate periodically, with the secondary mode trailing the primary mode by a fixed phase offset, a hallmark of nonlinear energy exchange through quadratic coupling. The stable, repeating peak structure confirms that the system has settled onto a limit-cycle attractor. Panel (b) shows the corresponding Fourier spectra in decibels. Both $|\psi_1|^2$ and $|\psi_2|^2$ display a cluster of sharp, evenly spaced lines symmetric about zero frequency, with the central peak at $0$ marking the dominant oscillation and the sidebands reflecting higher harmonics. The near-identical spectral envelopes of the two modes confirm that the comb is a collective property of the coupled system rather than a feature of either mode alone, and the well-resolved lines indicate a stable, phase-coherent limit cycle.

Figure~\ref{fig2} characterizes the modal intensities as functions of drive detuning $\Delta_1$. Panels (a)-(d) correspond to the primary mode $\psi_1$ and panels (e)-(h) to the subharmonic mode $\psi_2$. The mean intensities in panels (a) and (e) vary smoothly with weak non-monotonicity, reflecting a gradual energy redistribution away from resonance, with a shallow maximum near $\Delta_1 \approx 4.9$--$5.0$ suggesting a mild optimal resonance condition. The peak-to-peak amplitudes in panels (b) and (f) decrease monotonically with $\Delta_1$, consistent with weaker effective driving as the resonance condition is relaxed. The oscillation extrema in panels (c), (d), (g), and (h) follow the same trend -- both minima and maxima decrease smoothly, indicating a uniform contraction of the limit-cycle envelope without any distortion. Taken together, all eight panels show that $\Delta_1$ acts as a smooth control parameter that continuously adjusts the oscillation amplitude without triggering any qualitative change in dynamics, confirming the structural stability of the frequency comb state under drive detuning.

Figure~\ref{fig3} reveals a markedly different response under variation of $\Delta_2$, with a sharp transition that separates different dynamical regimes. For the primary mode $\psi_1$, panels (a)-(d) show that the mean intensity develops a pronounced peak near a critical $\Delta_2 \approx -4.5$, accompanied by strong amplification of the peak-to-peak amplitude in panel (b). The extrema tell the same story: the minimum approaches near-zero around $\Delta_2 \approx -5$, while the maximum spikes sharply, indicating a strongly distorted oscillation envelope consistent with the onset of efficient parametric coupling. The secondary mode in panels (e)-(h) provides direct evidence of the resulting energy transfer -- the mean intensity rises steadily before dropping abruptly at the same critical point where $\psi_1$ peaks, and the peak-to-peak amplitude in panel (f) follows suit. Most strikingly, the extrema in panels (g) and (h) display pronounced discontinuities, pointing to a transition between distinct limit-cycle configurations near $\Delta_2 \sim -4.5$ rather than a smooth continuous evolution.

Figure~\ref{fig4} shows how the modal intensity statistics evolve with the relative damping $\gamma_{21}$. For $\psi_1$, panels (a)--(d) present a consistent picture: the mean intensity grows monotonically with small superimposed combs emerging for $\gamma_{21} \gtrsim 0.5$, the peak-to-peak amplitude rises nearly linearly, and the extrema diverge -- the minimum falling toward near-zero while the maximum climbs -- reflecting a progressive widening of the comb envelope. The secondary mode in panels (e)-(h) broadly mirrors this, though the mean intensity in panel (e) exhibits non-monotonic combs at intermediate $\gamma_{21} \approx 0.4$--$1.2$, likely reflecting interference between competing energy-transfer pathways. Nevertheless, the peak-to-peak amplitude and extrema follow the same monotonic trends as $\psi_1$, and no discontinuities appear anywhere in the domain, confirming that coupling amplifies the comb continuously without triggering bifurcation.

Figure~\ref{fig5} maps the modal intensity statistics across the $(\Delta_1, f)$ parameter plane. The dominant feature in all panels is a strong gradient along the $f$ axis, which confirms that the amplitude of the drive is the primary control parameter. For both modes, the mean, peak-to-peak and extrema intensities increase smoothly with $f$ in nearly parallel contour bands, with $\Delta_1$ introducing only a slight tilt -- consistent with the secondary role already established in Fig.~\ref{fig2}. The most structurally informative feature appears in the minimum intensity panels (c) and (g), which reveal a threshold near $f \approx 15$-$20$: below it the minima remain near zero, above it they rise sharply, marking the onset of a fully developed limit cycle. No sharp boundaries or discontinuities appear elsewhere in the explored domain, confirming that the system remains on a single stable limit cycle throughout and that drive amplitude, not detuning, sets the fundamental energy scale of the oscillatory state.

Figure~\ref{fig6} maps the modal intensity statistics across the $(\Delta_2, f)$ parameter plane. For the primary mode $\psi_1$, panels (a)-(d) reveal a structured phase landscape shaped by a sharp resonance boundary near $\Delta_2 \approx -5.0$ to $-4.7$. Below this threshold, the mean intensity and peak-to-peak amplitude remain suppressed across all $f$, while above it a pronounced high-intensity wedge emerges, particularly for $f \gtrsim 25$, indicating strong parametric amplification. Panel (c) shows the minimum intensity that remains near zero for $f \lesssim 15$--$18$, forming a distinct boundary between weak combs and a fully developed limit cycle. This transition persists robustly throughout the $f$ range, pointing to a resonance-induced restructuring of the dynamical state. The secondary mode $\psi_2$ in panels (e)-(h) exhibits even sharper sensitivity to the same threshold. The mean intensity and peak-to-peak amplitude grow smoothly with $f$ for $\Delta_2 \lesssim -5.5$, but undergo rapid enhancement as $\Delta_2 \to -5.0$, with a distinct change in gradient near $\Delta_2 \approx -4.7$. Most strikingly, the minimum intensity in panel (g) shows a discontinuous jump from near-zero to finite values for $f \gtrsim 18$-$20$, and the maximum intensity in panel (h) peaks in the same high-amplitude wedge seen in $\psi_1$. The consistent localization of all these features around $\Delta_2 \sim -5$ establishes the secondary detuning as the principal gateway to the strongly nonlinear regime, where both modes undergo synchronized amplification and a qualitative transformation of the structure of the limit-cycle.

Figure~\ref{fig7} maps the modal intensity statistics across the $(\gamma_{21}, f)$ parameter plane. For the primary mode $\psi_1$, panels (a)-(d) show that the mean intensity, peak-to-peak amplitude, and maximum all increase monotonically with $f$, with a clear additional enhancement from coupling strength, particularly for $\gamma_{21} \gtrsim 1$. Panel (c) stands out: the minimum intensity remains near zero for $f \lesssim 15$-$18$, with a well-defined boundary above which it rises abruptly, marking the onset of fully sustained limit-cycle dynamics. This threshold shifts only weakly with $\gamma_{21}$, indicating that while coupling enhances the comb line amplitude, the transition to robust combs is primarily governed by drive strength. The secondary mode in panels (e)-(h) tells a broadly similar story, though with a notably weaker dependence on $\gamma_{21}$. The mean intensity, peak-to-peak amplitude and maximum all grow steadily with $f$, and panel (g) shows an analogous low-$f$ boundary where the minimum stays near zero before continuously rising. Crucially, no discontinuities or resonance wedges appear anywhere in this plane -- the system instead undergoes a smooth crossover from weak to strong oscillatory behavior. This contrasts with the sharp transitions seen in the $\Delta_2$ scans, and confirms that $\gamma_{21}$ modulates the efficiency of intermodal energy redistribution while drive amplitude remains the primary control parameter for the global dynamical state.

Figure~\ref{fig8} maps the absolute sensitivity of key intensity measures to infinitesimal variations in $\Delta_1$ across the $(\Delta_1, f)$ plane, complementing the amplitude maps of Fig.~\ref{fig5} by exposing where the system is most susceptible to parameter fluctuations. For both modes, the sensitivity is sharply localized along a narrow band near $f \sim 16$--$21$, coinciding with the oscillation onset boundary identified earlier. For $\psi_1$, panels (a) and (b) show that the mean intensity develops high-sensitivity patches at $f \gtrsim 25$, while the peak-to-peak amplitude is most responsive precisely along the transition boundary, where small changes in $\Delta_1$ can trigger large amplitude variations. The extrema in panels (c) and (d) remain nearly flat away from this band, with white regions marking parameter combinations where combs are absent and the relevant observables become ill-defined. The secondary mode in panels (e)-(h) displays a qualitatively similar but more pronounced structure, with larger sensitivity magnitudes reflecting stronger nonlinear responsiveness. Panel (e) shows broad regions of elevated sensitivity in large $f$, with irregular fluctuating patches pointing to complex intermodal energy exchange. The peak-to-peak amplitude in panel (f) again peaks along the same $f \sim 16$--$21$ boundary, and the extrema in panels (g) and (h) show enhanced gradients in the same zone, though with a smoother profile than $\psi_1$. Away from this band, the sensitivity drops markedly in both low- and high-amplitude regimes, confirming structural stability there. Together, the narrow high-sensitivity ridges aligned with the comb threshold indicate that the transition is governed by a sharply defined resonance condition. Once past this boundary, the system settles into a comparatively stable nonlinear regime, where detuning variations produce only moderate changes in the oscillatory state.

Figure~\ref{fig9} maps the absolute sensitivity of key intensity measures to variations in $\Delta_2$, directly probing the parametric resonance mechanism. In contrast to the diffuse sensitivity patterns seen for $\Delta_1$ in Fig.~\ref{fig8}, the response here is sharply localized along a well-defined resonance manifold. For the primary mode $\psi_1$, the sensitivity concentrates along a narrow diagonal ridge spanning $\Delta_2 \sim -5.2$ to $-4.5$ as $f$ increases from $\sim 20$ to $40$. Panel (a) shows the mean intensity that increases strongly along this ridge relative to the surrounding parameter space, marking a critical resonance condition. Panel (b) reveals an even stronger amplification in the peak-to-peak sensitivity, where a sharp filament-like structure signals that the oscillation amplitude is highly sensitive to small $\Delta_2$ variations in this regime. Panels (c) and (d) confirm that both the minimum and maximum intensities undergo abrupt changes along the same locus, though the minimum remains comparatively less sensitive. Outside this narrow band, uniformly low sensitivity confirms that the system is dynamically robust to $\Delta_2$ perturbations, and white regions again mark parameter combinations where sustained oscillations are absent. The secondary mode $\psi_2$ in panels (e)-(h) exhibits an even more pronounced sensitivity profile, reflecting its direct involvement in the parametric resonance process. Panel (e) shows the mean intensity developing a broadened high-sensitivity region at larger $f$, with fine-scale oscillatory patches pointing to complex intermodal energy exchange near resonance. Panel (f) reveals that the peak-to-peak amplitude attains its highest sensitivity along the same ridge as $\psi_1$, but with greater magnitude, identifying $\psi_2$ as the dominant carrier of nonlinear amplification. Panels (g) and (h) show that both the minimum and maximum intensities exhibit sharp gradients across the resonance boundary, with the maximum in panel (h) reaching extremely large sensitivity values, confirming a strongly nonlinear amplification mechanism. The consistent alignment of high-sensitivity structures across all eight panels establishes $\Delta_2$ as the primary control parameter governing parametric instability, with the bright filamentary ridge marking the precise resonance locus beyond which the system transitions abruptly between qualitatively distinct dynamical regimes.

Figure~\ref{fig10} maps the absolute sensitivity of key intensity measures to variations in $\gamma_{21}$, isolating the role of nonlinear energy exchange in shaping the steady-state dynamics. Unlike the sharply localized resonance structures seen in the $\Delta_2$ sensitivity maps, the response here is more distributed, with structure concentrated near a low-$\gamma_{21}$ boundary layer and in the strongly driven regime. For the primary mode $\psi_1$, panels (a)-(d) reveal two dominant features. Panel (a) shows fine-scale fluctuations in the mean intensity sensitivity for $f \gtrsim 20$ across a wide range of $\gamma_{21}$, indicating that in the strongly nonlinear regime small coupling variations lead to irregular modulations in energy distribution. Panel (b) shows the peak-to-peak amplitude is most sensitive near $\gamma_{21} \lesssim 0.2$ and $f \sim 12$-$18$, where a sharp gradient marks the transition from weakly coupled to dynamically active states. Panels (c) and (d) confirm elevated sensitivity in the same low-$\gamma_{21}$ region for both the minimum and maximum intensities, followed by a rapid decay to uniformly low sensitivity as coupling increases. White regions again correspond to parameter combinations without sustained oscillations. Once coupling is sufficiently strong, the primary mode becomes largely insensitive to further variations in $\gamma_{21}$. The secondary mode in panels (e)-(h) exhibits a richer sensitivity structure, reflecting its direct dependence on intermodal energy transfer. Panel (e) shows highly oscillatory sensitivity patterns at $f \gtrsim 25$ spanning a broad range of $\gamma_{21}$, suggesting complex interference effects in the energy exchange process. Panel (f) shows the peak-to-peak amplitude most sensitive at small $\gamma_{21}$ and low $f$, transitioning smoothly into a regime of moderate, nearly uniform sensitivity at larger drive. Panels (g) and (h) follow the same trend -- strong sensitivity near $\gamma_{21} \lesssim 0.2$ and $f \sim 12$-$18$, decaying gradually as coupling grows. Crucially, no sharp resonance ridge appears anywhere in this plane; instead, the sensitivity redistributes continuously across the parameter space, contrasting with the filamentary structures seen in Fig.~\ref{fig9} and confirming that $\gamma_{21}$ governs a smooth amplification process rather than a resonance-driven instability.

Now, we investigate the influence of drive power fluctuations on PFC-based sensors, where the frequency-comb parameters are expected to exhibit substantially lower sensitivity to variations in drive amplitude than to changes in the intrinsic system parameters. This inherent robustness is anticipated to improve sensing reliability under fluctuating excitation conditions. Figure~\ref{fig11} maps the absolute sensitivity of key intensity measures to variations in the driving amplitude $f$ across the $(\Delta_1, f)$ plane. Unlike the detuning sensitivities, which produced sharply localized resonance ridges, the response to $f$ combines a pronounced threshold effect with extended regions of gradual variation, consistent with its role as the global control parameter governing oscillation onset and growth. For the primary mode $\psi_1$, the sensitivity concentrates along a well-defined lower boundary near $f \sim 15$--$20$, with a slight upward shift as $\Delta_1$ increases from $4.5$ to $6.0$. Panels (b) and (c) show this boundary most clearly as a narrow high-intensity ridge, indicating that both the peak-to-peak amplitude and minimum intensity are highly sensitive to small forcing variations near the excitation threshold, consistent with a bifurcation-like entry into the limit-cycle regime. Panel (a) shows that above this boundary, for $f \gtrsim 22$, the mean intensity sensitivity becomes spatially distributed with comb patterns in $\Delta_1$, suggesting interference between nonlinear modes. Panel (d) shows that the maximum intensity sensitivity is comparatively smooth, gradually increasing with $f$ and $\Delta_1$ without sharp localized peaks except near the threshold. White regions again mark sub-threshold parameter combinations where oscillations are absent. The secondary mode in panels (e)-(h) displays qualitatively similar but quantitatively enhanced sensitivity. Panel (e) shows pronounced oscillatory bands extending across $\Delta_1$ for $f \gtrsim 20$, indicating that small forcing variations can trigger complex energy redistribution in the coupled system. Panel (f) confirms that the peak-to-peak amplitude is most sensitive along the same $f \sim 15$--$20$ boundary, with sensitivity gradually decaying above it. Panel (g) reveals that the minimum intensity is only weakly sensitive to $f$ away from the threshold, with a smooth gradient reflecting a gradual lifting of the oscillation floor. Panel (h) shows the maximum intensity sensitivity increasing monotonically with both $f$ and $\Delta_1$ without sharp discontinuities, indicating that the upper oscillation envelope is shaped by cumulative energy input rather than abrupt transitions. Taken together, the warm regions in all panels identify threshold-driven amplification zones, while the cooler regions confirm that once well above threshold, the system enters a saturated regime where further increases in $f$ produce only incremental dynamical changes.

Figure~\ref{fig12} maps the absolute sensitivity of key intensity measures to variations in $f$ across the $(\Delta_2, f)$ plane. Unlike the $(\Delta_1, f)$ case in Fig.~\ref{fig11}, the response here is strongly shaped by proximity to the parametric resonance condition, producing sharply localized sensitivity along a distinct resonance ridge rather than a simple threshold boundary. For the primary mode $\psi_1$, all panels are dominated by a narrow high-intensity ridge spanning $\Delta_2 \sim -5.2$ to $-4.5$ as $f$ increases from $\sim 20$ to $40$, where small changes in drive amplitude produce disproportionately large modal responses. Panel (a) shows oscillatory sensitivity patterns surrounding this ridge, reflecting interference effects and nonlinear modulation near resonance. Panels (b) and (d) confirm that the peak-to-peak and maximum intensities are most sharply enhanced along the same trajectory. Panel (c) stands apart: the minimum intensity is primarily sensitive along the lower onset boundary at $f \sim 14$--$18$, with only weak response near the resonance ridge. White regions again mark sub-threshold parameter combinations without sustained oscillations. The secondary mode in panels (e)-(h) displays an even more pronounced sensitivity landscape. Panel (e) reveals complex oscillatory bands surrounding the resonance ridge for $f \gtrsim 22$, indicating strong susceptibility of the mean intensity to small $f$ variations near $\Delta_2 \approx -5$. Panel (f) shows the peak-to-peak sensitivity reaching a sharply defined maximum along the same ridge, with values exceeding those of $\psi_1$ and underscoring the amplified nonlinear response of the secondary mode. Panel (g) shows the weakly sensitive minimum intensity except for a narrow band near the resonance boundary. Panel (h) shows the maximum intensity following a smooth but strongly increasing sensitivity profile along the resonance trajectory, peaking near $\Delta_2 \sim -4.5$ and $f \gtrsim 30$. Across all panels, warm regions identify zones of critical resonance-driven amplification, while cooler regions confirm dynamical saturation where further increases in $f$ produce diminishing sensitivity.

Figure~\ref{fig13} maps the absolute sensitivity of key intensity measures to variations in $f$ across the $(\gamma_{21}, f)$ plane, revealing how intermodal coupling modulates the system's responsiveness to external forcing. The structure combines threshold activation at low $f$ with coupling-mediated redistribution at larger drive, contrasting with both the diffuse patterns of Fig.~\ref{fig11} and the sharp resonance ridges of Fig.~\ref{fig12}. For the primary mode $\psi_1$, the sensitivity concentrates along a well-defined lower boundary near $f \sim 15$--$20$, with a weak upward shift as $\gamma_{21}$ increases. Panels (b) and (c) show this boundary most clearly as a sharp high-intensity ridge, where small forcing variations produce large changes in both the peak-to-peak amplitude and minimum intensity, consistent with a bifurcation-like entry into the oscillatory regime. Panel (a) shows that above this boundary for $f \gtrsim 20$, the mean intensity sensitivity develops oscillatory banded structures along $\gamma_{21}$, reflecting nonlinear interference induced by coupling. Panel (d) shows the maximum intensity sensitivity increasing smoothly and monotonically with both $f$ and $\gamma_{21}$, with no sharp features, indicating that the upper oscillation envelope responds gradually to increased forcing. White regions mark sub-threshold combinations where oscillations are absent.The secondary mode in panels (e)-(h) exhibits a richer sensitivity landscape. Panel (e) shows strong oscillatory structures across the accessible region for $f \gtrsim 20$, growing in complexity with $\gamma_{21}$ and pointing to enhanced nonlinear mixing at larger coupling. Panel (f) confirms that the peak-to-peak sensitivity peaks along the same threshold boundary $f \sim 16$--$20$ before gradually spreading across parameter space with increasing coupling. Panel (g) shows the weakly sensitive minimum intensity except near the activation boundary, where a narrow ridge marks the lifting of the oscillation floor. Panel (h) shows the maximum intensity sensitivity increasing smoothly with both $f$ and $\gamma_{21}$ without discontinuities, indicating that stronger coupling enhances the efficiency with which forcing amplifies the oscillatory state. As in the primary mode, warm regions identify threshold-dominated and coupling-enhanced amplification zones, while cooler regions confirm dynamical saturation well above threshold.

Figure~\ref{fig14} examines the oscillation frequency $\omega(\psi)$ of both modes as a function of detuning and coupling, complementing the intensity-based analysis by revealing how the temporal structure of the limit cycle responds to parameter variations. Panel (a) shows that as $\Delta_1$ varies across $[4.5, 6.0]$, the frequencies of both modes increase monotonically from $\omega \approx 1.38$ to $\omega \approx 1.405$ in near-perfect overlap, with no discontinuities. This robust frequency locking confirms that $\Delta_1$ induces only a gradual coherent shift in the collective oscillation frequency without disrupting synchronization, consistent with both modes behaving as a single effective oscillator throughout. Panel (b) tells a starkly different story under variation of $\Delta_2$. For $\Delta_2 \lesssim -5.7$, both modes track a common slowly rising frequency, but near $\Delta_2 \approx -5.7$ the primary mode $\psi_1$ jumps abruptly to a higher-frequency branch at $\omega \sim 3$, signaling a bifurcation or resonance-induced state change. The secondary mode $\psi_2$ follows a distinct trajectory, with delayed discontinuous jumps near $\Delta_2 \approx -5.0$ and $-3.8$, and a sharp dip near $\Delta_2 \approx -4.5$ pointing to a transient destabilization or switching event. Together these features confirm the central role of $\Delta_2$ in governing parametric resonance and inducing complex dynamical transitions. Panel (c) shows that for weak coupling $\gamma_{21} \lesssim 0.2$, the primary mode occupies a high-frequency state at $\omega \sim 3.3$, which collapses abruptly to $\omega \sim 1.5$ as coupling increases past this threshold. Beyond this transition, both modes converge and decrease gradually together across $\gamma_{21} \in [0.2, 2.0]$, reflecting the onset of strong synchronization where intermodal coupling enforces frequency locking and suppresses independent oscillatory behavior. The smooth decay at higher coupling suggests that increasing nonlinear interaction effectively renormalizes the oscillation frequency toward a unified, stabilized dynamical state.

Figure~\ref{fig15} maps the oscillation frequencies $\omega_{\psi_1}$ and $\omega_{\psi_2}$ across three two-parameter planes, linking the amplitude-based transitions discussed previously with their temporal signatures. Panels (a) and (b) show the frequency dependence on $(\Delta_1, f)$. Both modes increase smoothly with $f$, producing gradual blue-to-red contour bands in the vertical direction, while the dependence on $\Delta_1$ is weak enough to leave the contours nearly horizontal. A distinct stepped boundary near $f \sim 28$--$30$ separates a low-frequency regime from a high-frequency one, consistent with the oscillation onset identified in earlier figures. The close similarity between panels (a) and (b) confirms robust frequency locking across this parameter range. Panels (c) and (d) reveal a markedly different structure in the $(\Delta_2, f)$ plane. A sharply defined slanted ridge extends from $(\Delta_2 \sim -5.5,\ f \sim 20)$ to $(\Delta_2 \sim -4.2,\ f \sim 40)$, coinciding with the parametric resonance condition and marking a regime of rapid frequency increase. Both modes share nearly identical ridge structures, although panel (d) shows a slightly sharper localization, reflecting the enhanced responsiveness of $\psi_2$ near resonance. White regions at lower $f$ again correspond to sub-threshold domains without stable oscillatory solutions. Panels (e) and (f) show the dependence on $(\gamma_{21}, f)$. For $\psi_1$, panel (e) shows a rapid frequency increase with $f$ at low coupling, followed by a plateau-like transition around $\gamma_{21} \sim 0.5$--$1.0$, beyond which the frequency becomes weakly dependent on coupling strength, indicating entry into a synchronized regime. The secondary mode in panel (f) shows a smoother and more gradual dependence, with continuously bending contours and no sharp discontinuities, suggesting that coupling primarily regularizes the temporal dynamics rather than inducing abrupt transitions. Across all panels, the emergence of ridges and plateaus delineates the boundaries between threshold activation, resonance locking, and coupling-induced synchronization.

Figure~\ref{fig16} maps the differential sensitivity of the oscillation frequencies to the variations of the parameters, complementing the analysis of the amplitude sensitivity by identifying where the temporal dynamics are most susceptible to perturbations. The frequency sensitivities are markedly more localized than their amplitude counterparts, with sharp ridges tracing resonance boundaries and bifurcation loci. Panels (a) and (b) show the sensitivity of $\omega_{\psi_1}$ and $\omega_{\psi_2}$ to $\Delta_1$. Both panels are dominated by dark blue, confirming that the oscillation frequency is largely insensitive to primary detuning across the entire parameter space. Only a few isolated hotspots appear near $f \sim 28$--$32$ and $\Delta_1 \approx 4.8$--$5.4$, indicating minor resonance-induced perturbations. The sparse, fragmented nature of these features reinforces the conclusion from Fig.~\ref{fig15} that $\Delta_1$ plays no significant role in governing the temporal structure of the oscillations. Panels (c) and (d) reveal a starkly different picture for the sensitivity to $\Delta_2$. Panel (c) shows a diagonal sequence of localized peaks extending from $(\Delta_2 \sim -7.2,\ f \sim 34)$ to $(\Delta_2 \sim -5.2,\ f \sim 18)$, with a more pronounced ridge near $\Delta_2 \approx -5$ for $f \gtrsim 20$ marking the primary resonance transition. Panel (d) shows an even stronger and more continuous ridge for $\omega_{\psi_2}$, extending from $(\Delta_2 \sim -6.2,\ f \sim 40)$ to $(\Delta_2 \sim -4.5,\ f \sim 20)$, where the sensitivity reaches its maximum. These structures coincide with the frequency discontinuities identified in Fig.~\ref{fig14}, pointing to rapid frequency reconfiguration likely associated with mode switching or loss of phase locking. White regions at low $f$ again mark sub-threshold domains. Panels (e) and (f) show the sensitivity to $\gamma_{21}$. For $\psi_1$, panel (e) shows a curved band of enhanced sensitivity at $\gamma_{21} \lesssim 1$ and $f \sim 20$--$30$, aligned with the coupling-induced synchronization transition, beyond which the sensitivity rapidly diminishes as the frequency stabilizes. Panel (f) shows $\omega_{\psi_2}$ largely insensitive to $\gamma_{21}$ across most of the space, with only weak features near the activation boundary. This asymmetry indicates that the primary mode drives the collective frequency, while the secondary mode adjusts passively once synchronization is established.

Figure~\ref{fig17} maps the differential sensitivity of the oscillation frequencies to variations in $f$, revealing how external forcing modulates the temporal dynamics near activation thresholds and resonance boundaries. Panels (a) and (b) show the sensitivity as a function of $(\Delta_1, f)$. Panel (a) shows a prominent horizontal band of high sensitivity for $\psi_1$ near $f \sim 27$--$30$, shifting slightly upward as $\Delta_1$ increases from $4.5$ to $6.0$, marking a threshold regime where small forcing variations induce noticeable frequency changes. Above this band the sensitivity becomes weaker and oscillatory; below it the response is largely suppressed. Panel (b) shows $\psi_2$ is considerably more robust, with only a faint ridge near the activation boundary at $f \sim 16$--$20$, indicating that the secondary mode frequency is comparatively insensitive to direct forcing variations. Panels (c) and (d) show the sensitivity as a function of $(\Delta_2, f)$. Panel (c) shows a diagonal sequence of moderate sensitivity extending from $(\Delta_2 \sim -7.2,\ f \sim 34)$ to $(\Delta_2 \sim -5.2,\ f \sim 18)$, with a more intense vertical ridge near $\Delta_2 \approx -5$ for $f \gtrsim 20$ where the frequency becomes highly responsive to forcing. Panel (d) shows a stronger and more structured counterpart for $\psi_2$, with peak sensitivity concentrated along a curved ridge spanning $\Delta_2 \sim -6$ to $-4.5$ and $f \sim 20$--$35$. Both panels align with the frequency discontinuities identified in Fig.~\ref{fig14}, confirming that drive amplitude plays a critical role in modulating frequency near resonance. Panels (e) and (f) show the sensitivity as a function of $(\gamma_{21}, f)$. Panel (e) shows a high sensitivity stepped band for $\psi_1$ near $f \sim 26$--$30$, which moves upward as $\gamma_{21}$ increases from $0$ to $1$ before saturation, indicating that the interplay between coupling and forcing defines a threshold region of enhanced frequency tunability. Panel (f) again shows $\psi_2$ weakly sensitive to $f$, with only a narrow region of moderate response near $f \sim 15$--$20$ and $\gamma_{21} \lesssim 0.5$. Across all panels, the asymmetry between the two modes persists: $\psi_1$ carries the dominant and more structured frequency response to forcing, while $\psi_2$ remains comparatively passive except near activation and resonance boundaries.

\section{Conclusion}

This work develops a unified framework for characterizing sensitivity in PFC systems across their full nonlinear parameter space. By analyzing both amplitude and frequency observables under systematic variation of $\Delta_1$, $\Delta_2$, $f$, and $\gamma_{21}$, we show that sensitivity is not a passive byproduct of the dynamics but a structured reflection of the underlying nonlinear mechanisms. The primary detuning $\Delta_1$ produces smooth and spatially extended responses that preserve the structural stability of the comb state, while the secondary detuning $\Delta_2$ acts as the principal driver of sharply localized resonance boundaries across which both the amplitude and frequency observables undergo abrupt transitions. The driving amplitude $f$ sets the global oscillation threshold and dominates the overall energy scale, with sensitivity concentrated near activation boundaries, whereas the relative damping $\gamma_{21}$ continuously redistributes energy between modes without inducing sharp bifurcations.

Across all parameter planes, high-sensitivity regions are confined to narrow manifolds associated with resonance, threshold activation, and synchronization transitions, while large portions of parameter space remain dynamically robust. The frequency analysis adds a complementary layer to this picture, with sensitivity sharply localized along resonance ridges and bifurcation loci, particularly under variations in $\Delta_2$ and $f$. A clear division of dynamical roles emerges between the two modes: $\psi_1$ governs temporal coherence through a dominant and structured frequency response, while $\psi_2$ mediates nonlinear amplification through enhanced amplitude sensitivity. The strong spatial alignment between amplitude-based and frequency-based sensitivity structures confirms that both originate from the same mechanisms -- autoparametric resonance, nonlinear saturation, and coupling-induced synchronization. Taken together, these results show that PFC systems actively organize their sensitivity through intrinsic nonlinear interactions, offering a principled basis for engineering tunable sensing platforms that exploit localized sensitivity while maintaining global robustness.


\begin{thebibliography}{00}

\bibitem{b1}
S.~T. Cundiff and J.~Ye,
``Colloquium: Femtosecond optical frequency combs,''
\emph{Rev. Mod. Phys.}, vol.~75, no.~1, p.~325, 2003.

\bibitem{b2}
T.~J. Kippenberg, R.~Holzwarth, and S.~A. Diddams,
``Microresonator-based optical frequency combs,''
\emph{Science}, vol.~332, no.~6029, pp.~555--559, 2011.

\bibitem{b3}
A.~Ganesan, C.~Do, and A.~Seshia,
``Phononic frequency comb via intrinsic three-wave mixing,''
\emph{Phys. Rev. Lett.}, vol.~118, no.~3, p.~033903, 2017.

\bibitem{b4}
L.~S. Cao, D.~X. Qi, R.~W. Peng, M.~Wang, and P.~Schmelcher,
``Phononic frequency combs through nonlinear resonances,''
\emph{Phys. Rev. Lett.}, vol.~112, no.~7, p.~075505, 2014.

\bibitem{b5}
D.~A. Czaplewski \emph{et al.},
``Bifurcation generated mechanical frequency comb,''
\emph{Phys. Rev. Lett.}, vol.~121, no.~24, p.~244302, 2018.

\bibitem{b6}
A.~Keskekler \emph{et al.},
``Symmetry-breaking-induced frequency combs in graphene resonators,''
\emph{Nano Lett.}, vol.~22, no.~15, pp.~6048--6054, 2022.

\bibitem{b7}
R.~L. Kubena \emph{et al.},
``Phononic comb generation in high-Q quartz resonators,''
\emph{Appl. Phys. Lett.}, vol.~116, no.~5, 2020.

\bibitem{b8}
Q.~Yang \emph{et al.},
``Asymmetric phononic frequency comb in a rhombic micromechanical resonator,''
\emph{Appl. Phys. Lett.}, vol.~118, no.~22, 2021.

\bibitem{b9}
G.~Gobat \emph{et al.},
``Frequency combs in a MEMS resonator featuring 1:2 internal resonance,''
\emph{Nonlinear Dyn.}, vol.~111, no.~4, pp.~2991--3017, 2023.

\bibitem{b10}
J.~Sun \emph{et al.},
``Generation and evolution of phononic frequency combs via coherent energy transfer
between mechanical modes,''
\emph{Phys. Rev. Appl.}, vol.~19, no.~1, p.~014031, 2023.

\bibitem{b11}
S.~Vashahri-Ghamsari \emph{et al.},
``Magnomechanical phonon laser beyond the steady state,''
\emph{Phys. Rev. A}, vol.~104, no.~3, p.~033511, 2021.

\bibitem{b12}
Y.~Xu \emph{et al.},
``Nonreciprocal phonon laser in a spinning microwave magnomechanical system,''
\emph{Phys. Rev. A}, vol.~103, no.~5, p.~053501, 2021.

\bibitem{b13}
X.~Han \emph{et al.},
``Superconducting cavity electromechanics: realization of an acoustic frequency
comb,''
\emph{Phys. Rev. Lett.}, vol.~129, no.~10, p.~107701, 2022.

\bibitem{b14}
H.~Lei \emph{et al.},
``Nonlinear rovibrational excitation of phononic frequency combs in molecules,''
\emph{Opt. Express}, vol.~33, no.~3, pp.~5396--5410, 2025.

\bibitem{b15}
R.~Liu and G.~Zhu,
``Phononic frequency combs in twisted bilayer van der Waals materials,''
\emph{J. Appl. Phys.}, vol.~138, no.~14, 2025.

\bibitem{b16}
A.~Rangwala \emph{et al.},
``Spontaneous phononic frequency combs in solid-state systems,'' 2026.

\bibitem{b17}
Y.~Zheng, S.~M.~H. Gangaraj, M.~Park, J.~Wang, and A.~Ansari,
``Phononic frequency combs with tunable frequency spacing for ultrasensitive
temperature sensing,''
\emph{Phys. Rev. Appl.}, vol.~25, no.~2, p.~024022, 2026.

\bibitem{b18}
H.~Wu \emph{et al.},
``Precise underwater distance measurement by dual acoustic frequency combs,''
\emph{Ann. Phys.}, vol.~531, no.~9, 2019.

\bibitem{b19}
Z.~Qian \emph{et al.},
``Quadrature acoustic frequency combs multiplexing for underwater communications,''
\emph{Ann. Phys.}, vol.~534, no.~2, 2022.

\bibitem{b20}
L.~Bu, E.~Arroyo, and A.~A. Seshia,
``Frequency combs for MEMS vibration energy harvesters,''
in \emph{Proc.\ Transducers}, 2021, pp.~136--139.

\bibitem{b21}
T.~Zhang and A.~A. Seshia,
``A MEMS frequency comb energy harvester,''
\emph{J. Microelectromech. Syst.}, 2023.

\bibitem{b22}
R.~L. Kubena and W.~S. Wall,
``Enhanced stability oscillators using a phononic comb,''
U.S. Patent 11\,156\,897, 2021.

\bibitem{b23}
R.~L. Kubena and W.~S. Wall,
``Noise suppression in a phononic comb,''
U.S. Patent 11\,431\,293, 2022.

\bibitem{b24}
R.~L. Kubena \emph{et al.},
``Comb enhanced oscillator with AM-to-PM noise suppression,''
U.S. Patent 11\,606\,098, 2023.

\bibitem{b25}
R.~L. Kubena and W.~S. Wall,
``Phononic comb enhanced gradiometers,''
U.S. Patent 11\,567\,147, 2023.

\bibitem{b26}
R.~L. Kubena and L.~D. Sorenson,
``Phononic comb enhanced MEMS gravity gradiometers,''
U.S. Patent 11\,747\,512, 2023.

\bibitem{b27}
H.~J. Song, W.~S. Wall, and R.~L. Kubena,
``Phononic comb enhanced capacitive inductive sensor,''
U.S. Patent 11\,575\,348, 2023.

\bibitem{b28}
Z.~Qi, C.~R. Menyuk, J.~J. Gorman, and A.~Ganesan,
``Existence conditions for phononic frequency combs,''
\emph{Appl. Phys. Lett.}, vol.~117, no.~18, 2020.

\bibitem{b29}
A.~Ganesan, A.~Seshia, and J.~J. Gorman,
``Phononic frequency combs for engineering MEMS/NEMS devices with tunable
sensitivity,''
in \emph{Proc.\ IEEE SENSORS}, Oct.~2019, pp.~1--4.

\bibitem{b30}
A.~Ganesan,
``Transition from negative to positive parametric dependence in phononic frequency
combs,''
in \emph{Proc.\ 2024 Int.\ Conf.\ Modeling, Simulation \& Intelligent Computing
(MoSICom)}, Dec.~2024, pp.~396--399.

\bibitem{b31}
H.~M.~E. Hussein, S.~Kim, M.~Rinaldi, A.~Al\`{u}, and C.~Cassella,
``Passive frequency comb generation at radiofrequency for ranging applications,''
\emph{Nat. Commun.}, vol.~15, p.~2844, Apr.~2024.

\bibitem{b32}
S.~Marti, U.~von~L\"{u}pke, O.~Joshi, Y.~Yang, M.~Bild, A.~Omahen, Y.~Chu,
and M.~Fadel,
``Quantum squeezing in a nonlinear mechanical oscillator,''
\emph{Nat. Phys.}, vol.~20, no.~9, pp.~1448--1453, 2024.

\bibitem{b33}
M.~Mishra and A.~Ganesan,
``Tunable amplitude--frequency response of phononic frequency combs for quantum
sensing,''
in \emph{Proc.\ IEEE APSCON}, 2026, to be published in IEEE Xplore.

\end{thebibliography}
\end{document}